\input epsf
\input amstex
\loadbold
\documentstyle{amsppt}
\magnification=\magstep 1
\hsize29pc
\vsize42pc
\baselineskip=24truept
\def\mod{\text{mod}\,}
\def\res{\text{res}\,}
\def\d{\delta}
\def\psib{\bar{\psi}}
\def\psih{\hat{\psi}}
\def\psihb{\bar{\hat{\psi}}}
\def\g{\bold g}
\def\[{\left[}
\def\]{\right]}
\def\({\left(}
\def\){\right)}
\def\bpsi{\overline{\psi}}
\def\bl{\overline{\lambda}}
\def\l{\lambda}

\def\f{\bold f}
\def\gh{\hat{\bold g}}
\def\gt{\tilde{\g}}
\def\tr{\text{tr}\,}
\def\det{\text{det}\,}
\def\M{\Cal M}
\def\X{\Cal X}
\def\H{\Cal H}

\def\D{\Cal D}
\def\G{\Cal G}
\def\R{\Cal R}
\def\B{\Cal B}
\def\I{\Cal I}

\def\P{\Cal P}
\def\N{\Cal N}
\def\W{\Cal W}
\def\T{\Bbb S}
\def\Z{\Bbb Z}
\def\C{\Bbb C}
\def\m{\boldkey M}

\topmatter
\title An Additional Gibbs' State for the Cubic Schr\"{o}dinger Equation on the 
Circle\endtitle
\author K.L.  Vaninsky\endauthor
\affil Courant Institute\\
251 Mercer Street\\
New York University\\
New York, NY 10012 
\endaffil
\email vaninsky@math.ksu.edu\endemail
\thanks  The work is  partially supported by NSF grant DMS-9501002 and 
DMS-9971834\endthanks
\keywords  Spectral curve. Moduli. Symplectic volume\endkeywords
\subjclass 58F07, 70H15\endsubjclass
\abstract
An invariant Gibbs' state for the nonlinear Schr\"{o}\-dinger equation    
on the circle was constructed  by  Bourgain, \cite{B}, and  McKean, \cite{MC}, 
out of   the basic Hamiltonian   using a trigonometric cut-off. 
The cubic nonlinear Schr\"{o}dinger equation  is a completely integrable 
system having an  infinite number of additional integrals of motion. 
In this paper  we construct the second invariant Gibbs' state from one of 
these additional integrals for the cubic NLS on the circle. This additional 
Gibbs' state is singular with respect to the Gibbs' state previously 
constructed  from the basic Hamiltonian.
Our approach employs  the Ablowitz-Ladik system, a completely  integrable 
discretization  of the cubic Schr\"{o}dinger equation.
\endabstract

\endtopmatter
\rightheadtext{An additional Gibbs' State} 
\document
\subhead 1. Introduction \endsubhead The nonlinear Schr\"{o}dinger equation 
for the complex function $\psi(x,t),\; x\in \T,\; t\in R^1$ is 
$$
i\psi^{\bullet}=-\psi''+p|\psi|^{2p-2} \psi,
$$
where $p$ is an arbitrary integer.  It  can be written in Hamiltonian form. 
Let $\M=(\psi,\psib)$ be a space of pairs of  two complex functions
$\psi=Q+iP$ and $ \psib=Q-iP$ on the circle $\T$ of perimeter 1.
For any two functionals $F$ and $G$ on $\M$,  define
$$
\{F,G\}_{\omega}=i\int_{\T} {\d F\over \d \psi}{\d G\over \d
\psib} - {\d F\over \d \psib} {\d G\over \d \psi} dx.
$$
The Hamiltonian $\H(\psi,\psib)=\int_{\T} |\psi'|^2+ |\psi|^{2p} dx$ produces
the nonlinear Schr\"{o}\- - dinger flow
$$ 
\overset\bullet\to\psi=\{\psi, \H\}_{\omega} 
$$
The equation has two other generic integrals. They are $\N=\int |\psi|^2 dx$,  
the number of particles, and $\P=-\int i\psi'\psib \,dx$, the momentum.
 
The Gibbs' measure associated with the basic Hamiltonian
$$
e^{-{1\over 2} \H} d\, vol= e^{-{1\over 2} \H} {1\over \infty!}\,
\underset\infty\to\bigwedge\, \omega= e^{-{1\over 2} \H} {1\over \infty!}
\, \underset{x\in \T}\to\bigwedge \, i\d \psi(x) \wedge \d\psib(x)
$$
was constructed  by  Bourgain  and McKean \cite{B1-2, MC}; it 
is the product of two independent copies of circular Brownian motion for the  
components of the function $\psi=Q +iP$ coupled together by the
nonlinear factor $e^{-{1\over 2}\int_{\T}|Q^2 +P^2|^p dx}$. 
This measure is invariant under the flow;  for any $p$ the later exists 
almost everywhere with respect to the Gibbs' measure. 

The cubic NLS corresponds to the case $p=2$. It has infinite series of 
commuting Hamiltonians. The first five are listed below
$$
\align
\H_1& = \int|\psi|^2 dx,\\
\H_2& = \int -i \psi'\psib dx,\\
\H_3& = \int |\psi'|^2 + |\psi|^4 dx,\\
\H_4& = \int i \psi'''\psib-i\psib'\psi |\psi|^2-4 |\psi|^2 \psib \psi' dx,\\
\H_5&= \int |\psi''|^2 + 2 |\psi|^6 +8 |\psi'|^2|\psi|^2+\psib'{}^2\psi^2 +
\psi'{}^2 \psib^2 dx.
\endalign
$$
The classical generic integrals $\N,\;\; \P$ and $\H$ are the first three 
of this infinite series. All of them are integrals of the isobaric 
polynomials,  counting $\psi$ and $d/dx$ as having degree 1; for example, 
$\H_5$ is an integral of an isobaric polynomial of degree 6.

We consider  the phase space $\M$ of  pairs $(\psi,\psib)$ from
the Sobolev space $H^1$\footnote"*"{If $\psi(x)=\sum_k e^{2\pi i k x}
\hat{\psi}(k)$, then the Sobolev norm of $\psi(x)$ is $|\psi|^2_{H^s}=
\sum_k |\hat{\psi}(k)|^2 (1+k^2)^s$.}.  The original equation  defines a
global flow on $\M$, \cite{B2}. In this paper we construct  an invariant   
Gibbs' state from the higher Hamiltonian $\H_5$:
$$
e^{-{1\over 2} \H_5} d \, vol=
e^{-{1\over 2} \H_5} {1\over \infty!} \, \underset{x\in \T}\to\bigwedge\,
i\d \psi(x)\wedge \d \psib(x).
$$
This  state is  singular with respect to the  Gibbs' state
$e^{-{1\over 2}\H_3} d\, vol$.
To construct the measure and prove its invariance  we will use the
Ablowitz-Ladik (AL) system. 

The AL equation for the complex N-periodic function $\psi(n,t) \,, 
n\in Z^1,\, t \in R^1$,    
is\footnote"*"{$\bullet$  denotes a time derivative;
$\psi_n= \psi(n,t)$; periodicity means $\psi_{n+N}=\psi_n$ for $n\in \Z$.}
$$
i\overset\bullet\to\psi_n=- (\psi_{n+1} +\psi_{n-1} - 2\psi_n) +|\psi_n|^2
(\psi_{n+1} + \psi_{n-1}). 
$$
It is easy to check that the quantity
$$
D=\prod\limits_{k=0}^{N-1} R_k= \prod\limits_{k=0}^{N-1} (1-|\psi_k|^2)
$$
is an integral of motion. We  assume that $|\psi_n| < 1$
for all $n$. This area of the phase space we call the box $B$; it is invariant 
under the flow. In the box all quantities $R_n$ and $D$ are {\it positive}.

In fact, AL  has many integrals of motion. The first three  interesting ones are 
$I_0,\;I_2,\; I_4$: 
$$
\align
NI_0& \equiv  -{1\over 2} \log D ,  \\
NI_2&     =   \sum\limits_{r=0}^{N-1} \psi_r \bpsi_{r-1},\\
NI_4&     =\sum\limits_{r=0}^{N-1} \psi_r \bpsi_{r-2}\(1-|\psi_{r-1}|^2\) -
{1\over 2} (\psi_r \bpsi_{r-1})^2.
\endalign
$$
From these integrals we form   Hamiltonians
$$
\align
H_1& =N(I_0+\bar{I_0})\\
H_3& =N(I_2+\bar{I_2} - 2 I_0 -2\bar{I_0})\\
H_5& =N(I_4 +\bar{I_4} -4I_2- 4 \bar{I_2} +6I_0+6 \bar{I_0}).
\endalign
$$
Let $M_N$ be a space of complex $N$-periodic sequences $\psi_n= Q_n +iP_n,\; 
n \in Z^1$ with $|\psi_n| <1$.  Introducing  the bracket
$$
\{f,g\}_{\omega_0}\equiv i\sum\limits_{n=1}^{N} R_n\(
{\d f\over \d \psi_n} {\d g\over \d \psib_n}-
{\d f\over \d \psib_n} {\d g\over \d \psi_n}\),
$$
for two functionals $f$ and $g$ on $M_N$ 
we can write  the original AL  flow in  Hamiltonian form
$$
\overset\bullet\to\psi_n=\{ \psi_n, H_3\}_{\omega_0}. 
$$
The symplectic form $\omega_0$ producies the volume element
$$
d\, \text{vol}= {1\over N! D_N}  \underset{n}\to\bigwedge\;  i \d \psib_n\wedge \d
\psi_n.
$$
The volume of the box is infinite
$$
\int\limits_{B} d\, \text{vol} = \int\limits_{B}
{1\over N! D_N}\underset{n}\to\bigwedge\;i\d \psib_n\wedge \d\psi_n =\infty
$$
due to the singularity of the factor ${1\over D_N}$. 
On the box we can define the Gibbs' state with the density
$$
e^{-{N^5\over 2} H_5} d\, \text{vol}= e^{-{N^5\over 2} H_5} {1\over N!}
\, \underset{n}\to\bigwedge \, \omega_0 =
e^{-{N^5\over 2} H_5} {1\over N! D_N} \, \underset{n}\to\bigwedge\,
i \d\psi_n \wedge \d \psib_n, 
$$
The singularity of the volume element near the boundary is rectified  by the
vanishing factor  $ e^{-{N^5\over 2} 12 NI_0}= D^{3N^5}$:  the total mass
$$
\int\limits_{B}e^{-{N^5 \over 2}H_5 } d\, \text{vol} < \infty.
$$
The area of the box close to the boundary has negligible probability.  
The measure  of the area outside the box  we take  to be  zero.

Let $\epsilon={1\over N}$. Under the scaling $\psi_n=\epsilon 
\psi({n\over N})$ and $\theta=\epsilon^{-2}t$,   the AL vector field 
approaches the   NLS vector field as $N\rightarrow \infty$.  Indeed,
$$
\psi_{n+1} =\psi_n+\psi_n'\epsilon +{1\over  2} \psi_n'' \epsilon^2\hdots,
\quad \quad \quad
\psi_{n-1} =\psi_n - \psi_n'\epsilon +{1\over 2} \psi_n'' \epsilon^2\hdots.
$$
So,
$$
\psi_{n+1} +\psi_{n-1} -2\psi_n= \psi_n''\epsilon^2 + \hdots.
$$
and the flow in the time scale $\theta$, 
$$
i{\partial \psi_n\over \partial \theta}= -(\psi_{n+1}+\psi_{n-1} -2\psi_{n})
+ |\psi_n|^2(\psi_{n+1}+\psi_{n-1}),
$$
scales as follows:
$$
i\epsilon^3{\partial   \psi\over \partial  t} =-\epsilon
\psi''\epsilon^2 + \epsilon^3|\psi|^2 2\psi+\hdots.
$$
This indicates  that AL flow  converges to the NLS flow,   
though statement  can not be taken literaly because these two flows live  
in two different spaces $M_N$ and $\M$.

The integrals $H_1,\; H_3$ and $H_5$ of the AL system converge to $\H_1,\,
\H_3$ and $\H_5$:
$$
\align
& N\; H_1(\psi_n,\psib_n) \rightarrow \H_1(\psi,\psib),\\
-&N^3 H_3(\psi_n, \psib_n)\rightarrow  \H_3(\psi,\psib),\\
&N^5H_5(\psi_n,\psib_n)\rightarrow  \H_5(\psi,\psib).
\endalign
$$
Also, for any $\psi$,  we have $D_N(\psi_n,\psib_n) \rightarrow 1$,  as
$N\rightarrow\infty$.
Therefore, one should expect to get from the Gibbs' state for AL,  
in the limit of $N\rightarrow \infty$, the desired invariant measure
$$
e^{-{1\over 2}\H_5} d\, vol=
e^{-{1\over 2} \H_5}
{1\over \infty!}\, \underset{x\in \T}\to\bigwedge\,
i \d\psi(x)\wedge  \d\psib(x).
$$
Again, this statement is correct but  can not be taken literaly, because
these measures do not live on the same space.  
To make the arguments  rigorous we need to embed the AL 
flows and measures into the function space.  For this, we use 
interpolating trigonometrical polynomials. 

The main goal of the paper is the  space-time random field 
$\psi(x,t),$ $  x\in \T,\;t\in R^1$ such that:

(i) $\psi(x,t)$ is stationary respect $x$ and $t$;

(ii) $\psi(\bullet, t)$ has Gibbs' distribution $e^{-{1\over 2}\H_5}d\, vol$;

(iii) the random variable $\psi(\bullet,t),\;t\neq 0$ is measurable with 
respect to the $\sigma$-field generated by $\psi(\bullet,0)$;
the measure is supported by the solutions of NLS;

(iv) The $x$-derivative of  almost every realisation of the random field 
is H\"{o}lder continuous:
$$
|\partial_x \psi(x_1,t_1) -\partial_x \psi(x_2,t_2)| \leq K\[|x_1-
x_2|^{{1\over 2}-} +|t_1- t_2|^{{1\over 4}-}\]
$$
with a random constant $K$. The exponents ${1\over 2}$ and ${1\over 4}$ 
are optimal.

The paper is organised as follows. In the section 2 we introduce the commuting 
flows of the AL hierarchy and commuatator formulas for them. 
Section 3 is devoted to the study of the direct spectral problem for the 
auxiliary linear system. We explain the details  of the spectral curve  for 
various types of potentials. Invariant quantities are computed in section 4. 
The Floquet and dual Floquet solutions are studied in sections 5 and 6. 
Section 7 provides formulas for the symplectic structure and the Poisson 
bracket. This completes  the first part of the paper about integrability 
properties and Hamiltonian formalism  for the AL system.

The second part starts with  section 8 where the strategy for constructing the 
measure is outlined; it is implemented  in the subsequent sections 
9-11. 

Finally I would like to thank I. Krichever, H. McKean, V. Peller and J. Zubelli 
for helpful discussion. It is also  pleasure to thank MPI in Bonn and IMPA 
for their  hospitality.

\subhead 2. Ablowitz--Ladik hierarchy\endsubhead 
The Ablowitz--Ladik equation for the complex $N$--periodic function 
$\psi(n,t) \, n\in Z^1,\, t \in R^1$ is
$$
i\overset\bullet\to\psi_n=- (\psi_{n+1} +\psi_{n-1} - 2\psi_n) +|\psi_n|^2 
(\psi_{n+1} + \psi_{n-1}). \tag 1
$$ 
Recall that 
$$
D=\prod\limits_{k=0}^{N-1} R_k= \prod\limits_{k=0}^{N-1} (1-|\psi_k|^2)
$$
is an integral of motion, whence the "box" $B=(\psi:\; |\psi_n| < 1 
\;\text{for all}\;n)$ is invariant under the flow. In the box 
$R_n$ and $D$ are {\it positive}. 

The flow (1) is one of infinitely many flows of the AL hierarchy.
The rotation or the phase flow
$$
i\overset\bullet\to\psi_n=-  \psi_n \tag 2	
$$
is the first flow of the hierarchy. Let
$$
V_2(n,t,\l)= {1\over \sqrt{R_n}} 
\[ \matrix 
\lambda & \psi_n \\
\bpsi_n & \lambda^{-1} 
\endmatrix \].
$$
The phase flow is the compatibility condition for 
$$
\[\partial_t -V_1,\Delta- V_2\]=0,    \tag 3
$$
where $V_1=i\sigma_3/ 2$ and $\Delta f_n=f_{n+1}$ is a shift operator. 
Here and below  $\sigma$  denotes the  Pauli matricies
$$
\sigma_1=\left(\matrix
 0& 1\\
1 & 0
\endmatrix \right), \quad
\sigma_2=\left(\matrix
0 & -i\\
i &  0
\endmatrix \right), \quad
\sigma_3=\left(\matrix
1 & 0\\
0 & -1
\endmatrix \right).
$$

The original AL equation (1) is the compatibility condition 
for\footnote"*"{A similar form of the  
commutator formalism was considered in \cite{AL, MEKL}. Our form has the  
small advantage, that it leads to a {\it unimodular} monodromy matrix.} 
$$
\[\partial_t- V_3, \Delta- V_2\]=0,   \tag 4
$$
where
$$
\align
V_3& (n,t,\lambda) \\
& =i \[\matrix \lambda^2-1+{3\over 2} 
\psi_n \bpsi_{n-1}-{1\over 2}\bpsi_n\psi_{n-1} & 
\psi_n \lambda - \psi_{n-1} 
\lambda^{-1} \\
\bpsi_{n-1} \lambda -\bpsi_n \lambda^{-1} & 1  - \lambda^{-2} + {1\over 2}\bpsi_n \psi_{n-1} + 
{1\over 2}\psi_n \bpsi_{n-1}
\endmatrix \].
\endalign
$$
The formula means that\footnote"**"{The operator $V=V_2$ acts like shift: 
$V_2(n)\f_n=\f_{n+1}$.} 
$$
(\partial_t- V_3(n+1))(\Delta- V_2(n))-(\Delta- V_2(n))(\partial_t -V_3(n))=0
$$
i.e. 
$$
\overset\bullet\to V_2(n)=V_3(n+1)V_2(n) -V_2(n)V_3(n).
$$

\subhead 3. Monodromy matrix. The spectral curve \endsubhead
Since $\psi_{n+N}= \psi_n$, all matricies $V_k,\, k=1,2,\hdots$, satisfy the periodicity 
condition $V_k(n+N)=V_k(n)$. We  introduce  the transition matrix 
$$
T_{n,m}(t,\l) \equiv V(n-1,t,\l) V(n-2,t,\l)\dotso V(m,t,\l) \quad \quad 
(n>m).
$$
It is easy to see that the spectrum of $T_{m+N,m}(t,\l)$ does not depend on $m$. Indeed,
$$
T_{m+1+N,m+1}(t,\l)= V(m+N) \dotso V(m+1)= V(m+N) T_{m+N,m} V(m)^{-1}.
$$
Also, the spectrum of $T_N\equiv T_{N,0}$ does not depend on  time. This 
follows from the identity $\overset\bullet\to T_N =[ V_3(0), T_N]$, which 
is  proved as follows:

$$
\align
\partial_t& [V \dotso V(k+1)V(k) \dotso V] \\
&= \dotso +V\dotso  \overset\bullet\to  V(k+1)  V(k) \dotso V + 
V\dotso  V(k+1)   \overset\bullet\to V(k)  \dotso V+  \dotso  \\
&=\dotso + V\dotso (V_3(k+2) V(k+1)- V(k+1) V_3(k+1)) V(k) \dotso V \\  
& \quad \quad \quad +V\dotso  V(k+1)  (V_3(k+1) V(k)- V(k) V_3(k) ) \dotso V  + \dotso \\
& = V_3(N) T_N(\l)- T_N(\l) V_3(0).
\endalign
$$

Consider the special "Floquet" solution 
$$
\g_n=\[ \matrix g^1_n(\l)\\  g^2_n(\l) \endmatrix \]
$$
of the eigenvalue problem $\g_{n+1}(\l)=V(n,\l) \g_n(\l)$ specified  by the 
condition 
$$
\g_N(\l)=T_N \g_0(\l)=w \g_0(\l),
$$
where $w$ being the  complex "multiplier" determined by 
$$
0=\det \[\matrix T_N^{11} -w  & T_N^{12}\\
                   T_N^{21}   & T_N^{22}-w \endmatrix \] = w^2 -2 w \Delta(\l) +1,
$$
where $\Delta(\l)=\tr T_N(\l)/2$. We know that $\Delta(\l)$ 
is an integral of motion. The multiplier is 
$$
w = \Delta(\l) +\sqrt{\Delta^2(\l) -1 };
$$
It becomes single-valued on the spectral curve \break 
$\Gamma=\{Q=(\l, y)\in {\Bbb C}^2 :\;\; y^2 =\Delta^2(\l) -1\}$. 

The curve $\Gamma$  inherits  symmetries from $V_2(n,\l)$. 
It is easy to check that
$$
\align
\sigma_1 V\({1\over \bl}\) \sigma_1 & =\overline{V(\l)}, \tag 1 \\
\sigma_3 V(-\l) \sigma_3 & =-V(\l). \tag 2
\endalign
$$
Obviously $T_N(\l)$ satisfies similar identities.  Whence 
$$
\align 
\Delta\({1\over \bl} \) & = \overline{\Delta(\l)}, \tag 3 \\
\Delta(-\l)& = (-1)^N \Delta(\l). \tag 4
\endalign
$$ 
The symmetries (3-4) produce an antiholomorhic involution $\tau_a$ and 
a holomorphic involution $\tau$ on the curve  
$$
\align
& \tau_a:\quad \quad \quad  (\l,y) \longrightarrow \({1\over \bl},
\overline{y}\), 
\tag 5 \\
&\tau:\quad \quad \quad  (\l,y) \longrightarrow \(- \l,(-1)^N y\). 
\tag 6
\endalign
$$
On $\Gamma$, there  exists another holomorphic involution $\tau_{\pm}$ which 
permutes the sheets, viz. 
$
\tau_{\pm}:   (\l,y) \rightarrow \(\l , -y\). 
$
From the quadratic equation for $w$, we have \break $w(Q)w(\tau_{\pm}Q)=1$. 

\noindent{\bf Remark.} Obviously, we have  freedom in the choice of  sign 
for the second coordinate of the involution, say
$$
\tau_a: \quad \quad    (\l,y) \longrightarrow  \( {1\over \bl} , \pm 
\overline{y} \).
$$
The sign in (5--6) is chosen in such way that $\tau_a$ and $\tau$  preserve 
the infinities of the curve (see example 5 of this section).

\noindent{\bf Example 1:} vanishing  potential $\psi_n\equiv 0$. Then
$$
T_N(\l) =\( \matrix \l^N & 0 \\ 0 & \l^{-N} \endmatrix \),  \quad \quad 
\Delta(\l)= \cosh N \log \l \quad \text{and}\quad w(Q)= e^{\pm N \log \l(Q)}.
$$
The points $\l_{k}^{\pm}= e^{i{2\pi\over 2N} k },\, k=0,\cdots, 2N-1$ satisfy 
the condition $\Delta^2(\l_k^{\pm})=1$ and at these 
points\footnote"*"{$\bullet$ now denotes 
derivative in $\l$ variable.} $\Delta^{\bullet}(\l_k^{\pm})=0.$ The points 
$\l_k^{\pm}$ are simple crossings of the curve $\Gamma$. 
They form periodic/antiperiodic spectrum; 
namely, $w=+1$ for $k$ even and $w=-1$ for $k$ odd. If $\l =e^{i\theta}$, then 
$\Delta(e^{i\theta})=\(e^{i\theta N}+ e^{-i \theta N}\)/  2= \cos \theta N$. 
The graph of $\Delta(\l)$ for  $N=2$ is shown in fig 1.
\midinsert
\epsfxsize=250pt
\centerline{
\epsfbox{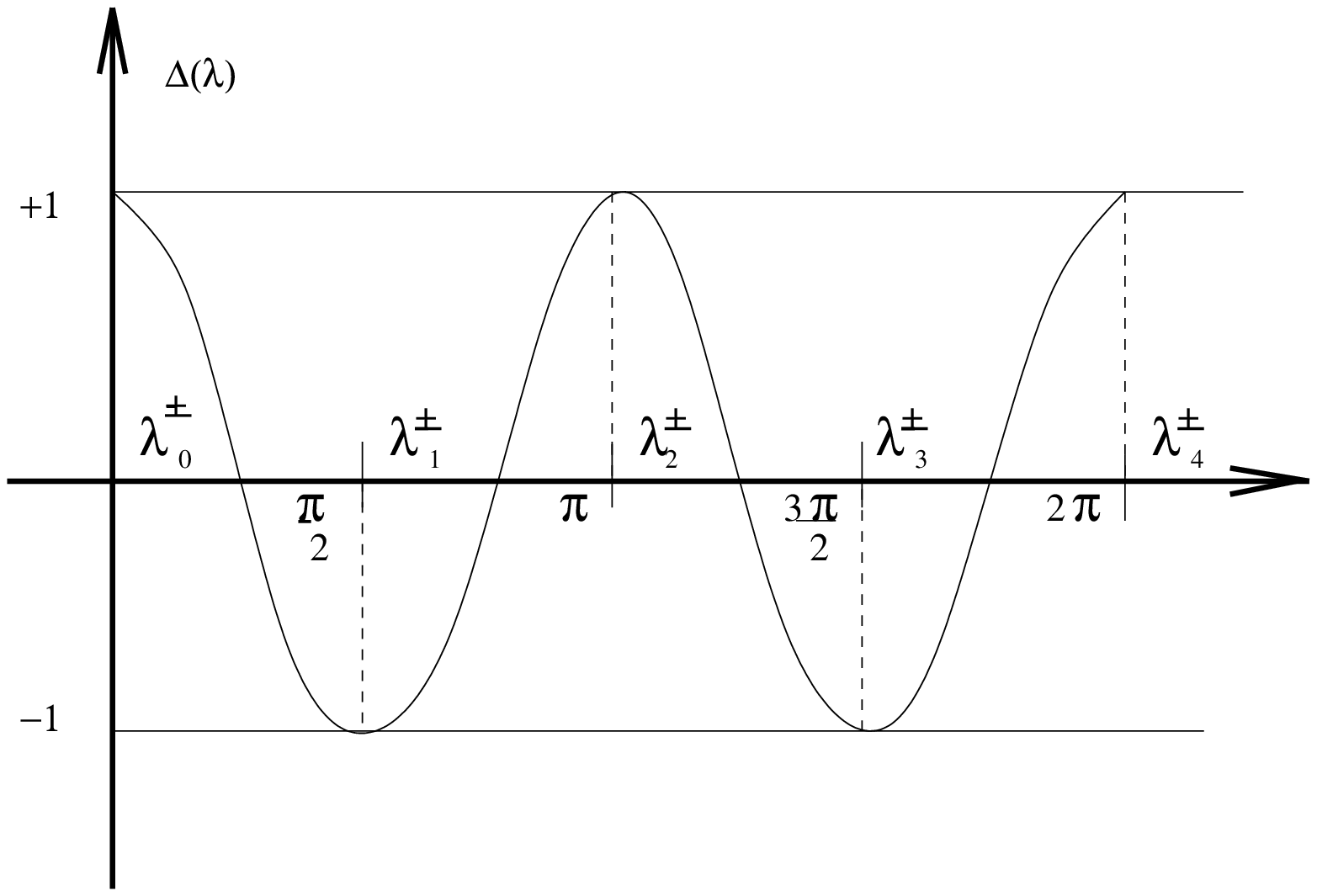}
}
\botcaption{fig. 1}
\endcaption
\endinsert

When $\psi_n$ is not identicaly zero,  then the double points $\l_{k}^{\pm}$ 
split into pairs $\l_{k}^{-}$ and  $\l_{k}^{+}$ of simple ramification. 
The symmetry (3) implies that $\Delta(\l)$ is always real for $|\l|=1$. Note 
also that,  due to the symmetry (4),  the branch points 
form symmetric pairs $\l_{k}^{-},\, \l_{k}^{+}$ and i
$\l_{k'}^{-},\, \l_{k'}^{+}$ such 
that $\l^{\pm}_k=- \l_{k'}^{\pm}$, when $k-k'\equiv 0$ (mod $N$). 

The branch points $\l_{k}^{\pm}$ lie only on the unit 
circle\footnote"**"{The statement 
is proved by the  method of \cite{AL}.}.  Indeed, the equation 
\break$\g_{n+1}(\l) = V(n,\l) \g_n(\l)$ 
can be written in the form $\Lambda_n \g_n(\l) =\l \g_n(\l)$, where
$$
\Lambda_n=  \[\matrix \sqrt{R_n}\Delta  & - \psi_n \\
                    \bpsi_{n-1} & \sqrt{R_{n-1}} \Delta^{-1}
\endmatrix \] \quad \quad \text{and} \quad  \Delta f_n= f_{n+1}.
$$ 
It is easy to see that $\Lambda$ is unitary in the space of 
$2N$--periodic vector functions  with the complex inner product 
$<\f,\g>_{2\heartsuit} =
{1\over 2N}\sum_{k=0}^{2N-1} f_k \overline{g}_k$. Indeed, introducing 
the formal inverse
$$
\Lambda^{-1}_n=\[ \matrix \sqrt{R_{n-1}} \Delta^{-1} & \psi_{n-1} \\
                       -\bpsi_n     & \sqrt{R_n} \Delta \endmatrix \]
$$
it is easy to derive the Cauchy identity
$$
\align
<\Lambda \f, \g >_{\heartsuit}  & = <\f, \Lambda^{-1} \g>_{\heartsuit} + \\
&+{1\over N} \sqrt{R_{N-1}} \( f^1_N \overline{g}_{N-1}^1 - f_0^1 \overline{g}_{-1}^1 \) +
{1\over N} \sqrt{R_{N-1}} \( f^2_{-1} \overline{g}_{0}^2 - f_{N-1}^2 \overline{g}_{N}^2 \),
\tag 7
\endalign 
$$
where $<\f, \g>_{\heartsuit} = {1\over N} \sum_{k=0}^{N-1} f_k \overline{g}_k$. 
Appling this formula twice, first to the interval $k=0,\cdots, N-1$ and then to the 
interval  $k=N,\cdots, 2N-1$  we obtain a similar identity for 
$<\bullet, \bullet>_{2\heartsuit}$ which implies the result.

\noindent{\bf Example 2:} $N=2$--periodic case.  It is instructive  to 
analyze this case complitely. The potential can be written in the form 
$$
\psi_n= \cases A+B, & n \;\; \text{is even}\\
               A-B, & n \;\; \text{is odd}. \endcases
$$
The matrix $T_2$ can be easily computed: 
$$
T_2= {1\over \sqrt{D}} \[\matrix \l^2 + ( A-B)(\overline{A}+ \overline{B}) & 
                         \l(A+B)+  \l^{-1}(A-B)\\
                         \l(\overline{A}- \overline{B}) + \l^{-1} 
(\overline{A}+ \overline{B}) &
            \l^{-2}+ (A+B) (\overline{A}- \overline{B})  \endmatrix\].
$$
Then $\Delta(\theta)= {1\over \sqrt{D}} \[ \cos 2\theta +|A|^2 -|B|^2\]$ for  
$\l=e^{i \theta}$,  and $D = (1-|A+B|^2)$ $\times(1-|A-B|^2)$.
Consider the case $B\equiv0$ and $A\neq 0$. Then $\Delta(\theta)= {1\over 
\sqrt{D}} \[ \cos 2\theta +|A|^2 \],$  and $D = (1-|A|^2)^2.$
The graph of $\Delta(\theta)$ is shown in fig. 2.
\midinsert
\epsfxsize=250pt
\centerline{
\epsfbox{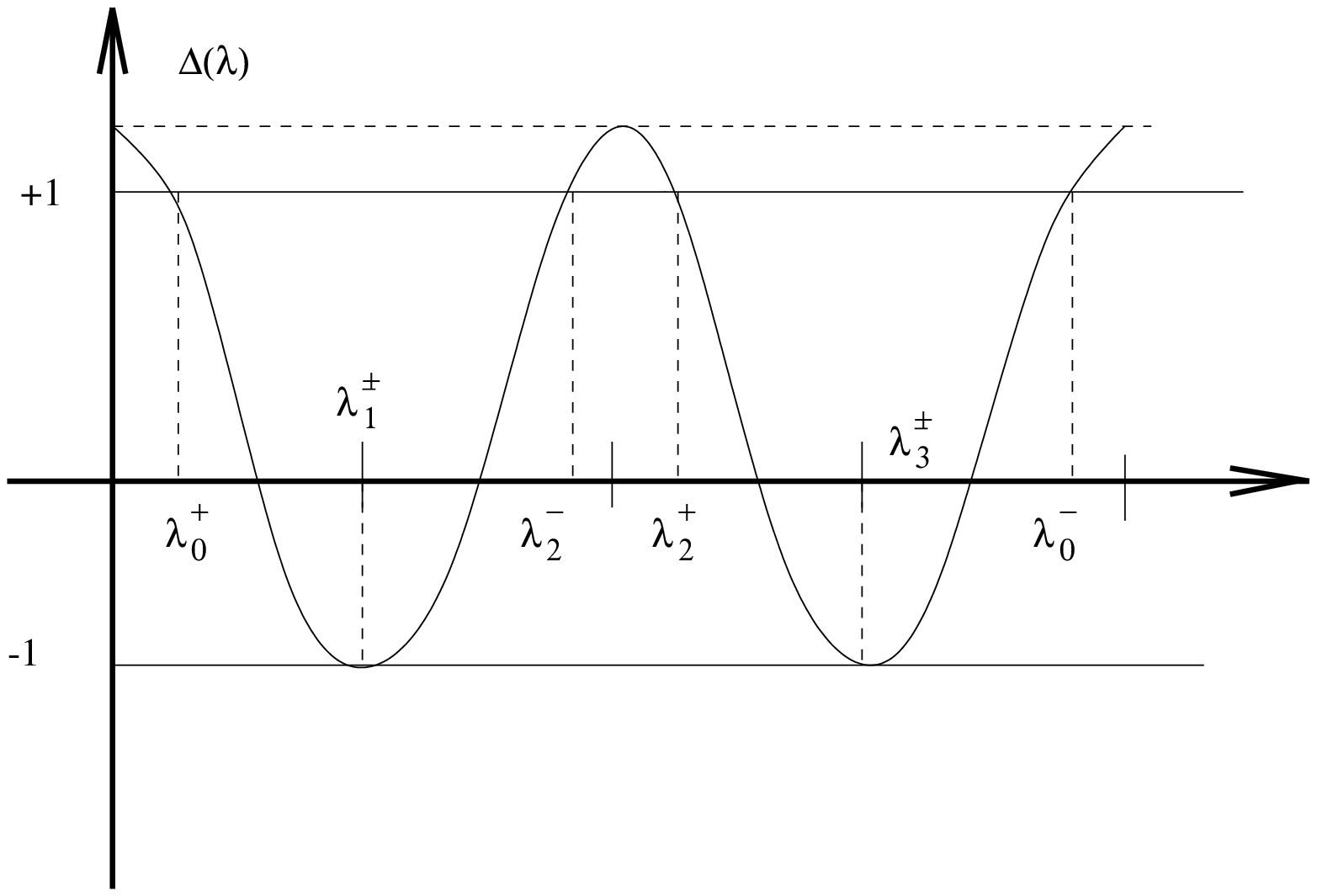}
}
\botcaption{fig. 2}
\endcaption
\endinsert
\noindent
The double points $\l_0^{\pm}$ and $\l_2^{\pm}$ split into pairs, while  
$\l_1^{\pm}$ and $\l_3^{\pm}$ remain double. 

Now consider the case when $A=0$ and $B\neq 0$. 
Then,  
$
\Delta(\theta)= {1\over \sqrt{D}} [ \cos 2\theta $ $  -|B|^2 ],$ $   D = (1-|B|^2)^2.
$
In this case,  the double points $\l_0^{\pm}$ and $\l_2^{\pm}$ do not split, 
while $\l_1^{\pm}$ and $\l_3^{\pm}$ split into pairs of simple roots.

Finally, there is an open area in the space of parameters, say, 
$A\sim B$, where 
$$
\Delta(\theta)\approx  {1\over \sqrt{D}}  \cos 2\theta , \quad \quad D \approx  1-4|A|^2 < 1.
$$
In this case {\it all} double roots split into  pairs of simple roots.

\noindent{\bf Example 3:} constant potential $\psi_n=\psi_0$. Consider, first, 
the case $N=2$. Then 
$$
T_2= {1\over \sqrt{D}} \[\matrix \l^2 + |\psi_0|^2 & \l\psi_0 +  
\l^{-1}\psi_0\\
\l \overline{\psi}_0 + \l^{-1} \overline{\psi}_0 &  \l^{-2} + 
|\psi_0|^2 \endmatrix\],
$$
$ \Delta(\theta)= {1\over \sqrt{D}} \[\cos 2\theta +|\psi_0|^2 \],$ 
and  $D = (1-|\psi_0|^2)^2$. 
The branch points $\l_k^{\pm}$ are given by the equation $\Delta^2(\l)=1$. 
Obviously, the points $\l^{\pm}_1$ and $\l^{\pm}_3$ do not split. For 
$\l_0^{\pm}$ we have the equation
$$
1-|\psi_0|^2=\cos 2\theta + |\psi_0|^2,
$$
so for small  $|\psi_0|$,   
$$
1- 2 |\psi_0|^2 = 1 - 2  \theta^2 +O(\theta^4).
$$
Finally, we arrive at  $\theta= |\psi_0| +O(|\psi_0|^3)$ and 
$$
\l_0^{\pm} =e^{\pm i (|\psi_0| +O(|\psi_0|^3))} \quad \quad \quad 
\l_2^{\pm} =e^{\pm i (|\psi_0| +O(|\psi_0|^3)) + i\pi}.
$$
In words, the open gap is proportional to the absolute value of the 
potential $|\psi_0|$. 

The case of general $N$ can be treated easily. One has to compute spectrum 
of $T_N$.  After simple algebra
$$
\l_0^{\pm} =e^{\pm i (|\psi_0| +O(|\psi_0|^3))},\quad\quad \quad 
\l_N^{\pm} =e^{\pm i (|\psi_0| +O(|\psi_0|^3)) + i\pi}.
$$

\noindent{\bf Example 4:} two-gap trigonometric potential. Consider 
the $N$--periodic potential $\psi^k$:
$$
\psi_n^k=e^{i\phi_k n} \psi_0,\quad\quad \quad \phi_k={2\pi k\over N}\quad   
k=0, \cdots, N-1.
$$
We will show that this potential opens $k$--th and $k+N$--th gap as in  
$$
\l_k^{\pm}=e^{i \phi_k/2 \pm i (|\psi_0| +O(|\psi_0|^3))}, \tag 8 
$$
and
$$
\l_{k+N}^{\pm} = e^{i \phi_k/2 \pm i (|\psi_0| +O(|\psi_0|^3)) + i\pi}. \tag 9
$$
To emphasise the dependence on $\psi^k$ we write 
$V(n,\l|\psi^k) = V^k(n),\; T(\l|\psi^k)=
T^k(\l)$ and $\Delta(\l| \psi^k)= \Delta^k(\l)$.  Note, first,  that
$$
L^k(n)=\Phi^n V^k(0) \Phi^{-n}, \quad \text{where} \quad \Phi=e^{i\sigma_3 
\phi_k/2}.
$$
Therefore,
$$\align
T(\l|e^{i\phi_k n}\psi_0) & = V^k(N-1) \cdots V^k(0) =  \Phi^N(\Phi^{-1} V^k(0))^N\\
& = (-1)^k (\Phi^{-1} V^k(0))^N= (-1)^k T (\l e^{-i\phi_k/2}|e^{-i\phi_k/2}\psi_0).
\endalign
$$ 
Taking the trace we obtain
$$
\Delta(\theta|  e^{i\phi_k n} \psi_0)= (-1)^k 
\Delta(\theta - {\phi_k\over 2}|  e^{- i\phi_k/2} \psi_0).
$$
This and the result of Example 3 produce the formula (8).  $(9)$ follows 
from this and the symmetry (4). 

\noindent{\bf Remark 1.} The situation here is similar  to the NLS equation, 
\cite{MCV1}, where the $N$--th gap opens in proportion to the  $N$-th Fourier 
coefficient.

\noindent{\bf Remark 2.} The $4N$ branch points 
$\l_0^{\pm}, \cdots, \l_{2N-1}^{\pm}$ determine  the curve. The symmetry (4) 
leaves only $2N$ free  parameters. In fact, there are only $N$ free 
independent parameters due to periodicity conditions.

\noindent{\bf Example 5:} generic periodic potential. The potential can 
be written as a sum of harmonics $\psi^k,\quad  k=0, \cdots , N-1$. All gaps 
are open.  To construct the curve $\Gamma$,  let us take two copies ``+'' 
and ``-'' of $\Bbb CP^1$ cut along circular arcs connecting $\l_k^-$ and 
$\l_k^+$ (fig. 3) \midinsert
\epsfxsize=250pt
\centerline{
\epsfbox{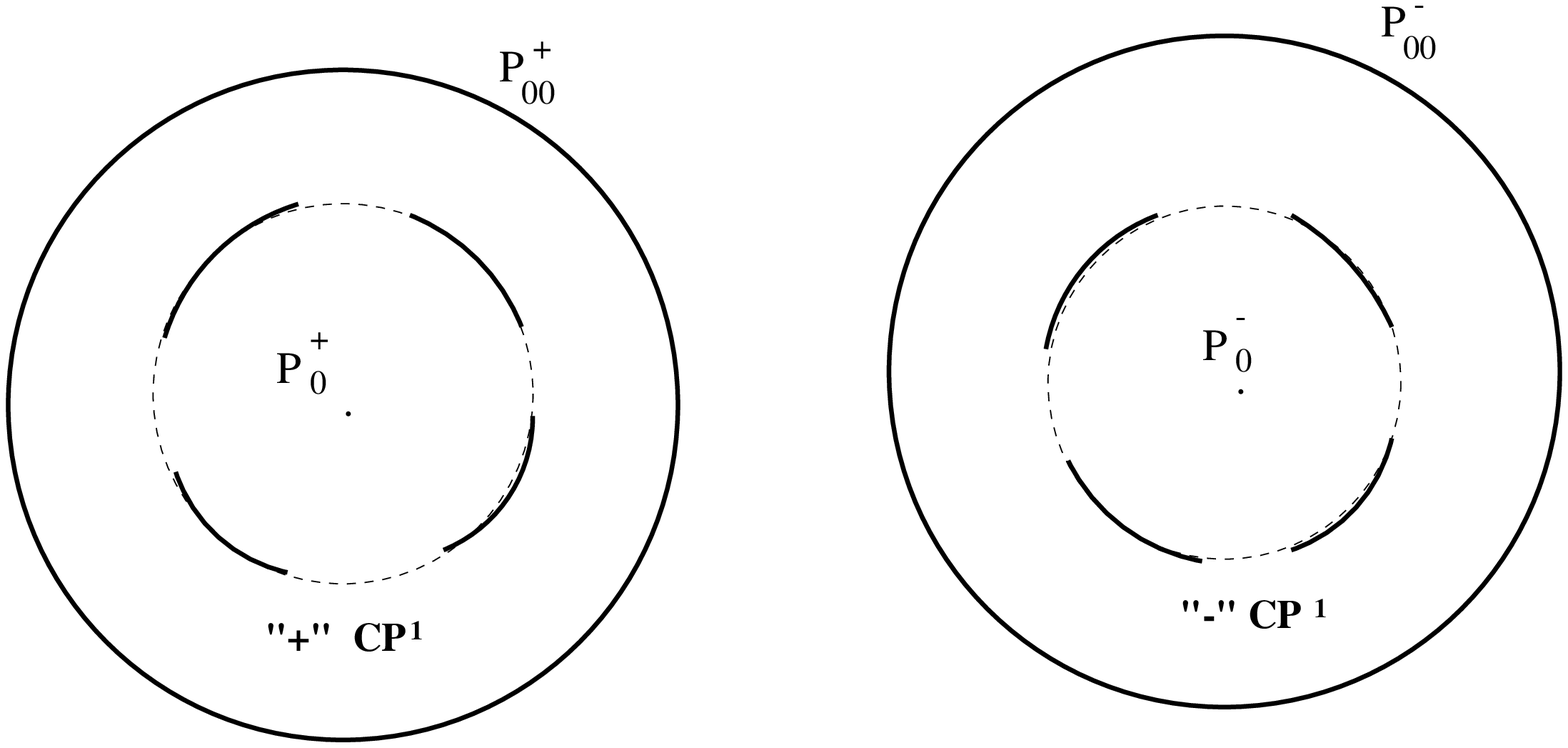}
}
\botcaption{fig. 3}
\endcaption
\endinsert

Each copy  $\Bbb CP^1$ has two marked points $P_{0/\infty}^+$ and 
$P_{0/\infty}^-$. The behavior of $\Delta(\l),\;\; y(Q)= 
\sqrt{\Delta(\l)^2-1},$ and $w(Q)$ near these points is this:
$$
\Delta(\l)\sim {1\over 2} \l^N,\; y(Q)\sim \pm {1\over 2} \l^N, \;  
w(Q) \sim \l^{\pm N},\quad \quad \quad \quad \l=\l(Q),\; Q\in (P_\infty^{\pm});
$$
and
$$
\Delta(\l)\sim {1\over 2} \l^{-N},\; y(Q)\sim \pm{1\over 2}\l^{-N}, \;  w(Q) \sim 
\l^{\mp N}, \quad\quad \quad \quad    \l=\l(Q),  \;
Q\in (P_0^{\pm}).
$$
From this, we see that $\tau_a$ maps  point $Q=(\l, y)$ in the vicinity 
of $P_{\infty}^{+/-}$ to the  point $\tau_a Q=({1\over \overline{\l}}, 
\overline{y})$ in the vicinity of $P_0^{+/-}$. The curve is 
obtained by gluing together ``$\pm$'' copies of $\Bbb CP^1$ along the cuts and identifying the points $P_{\infty}^{+/-}$ 
with $P_0^{+/-}$  as specified 
$$
P^+=P_{\infty}^++ P_0 ^+,\quad \quad \text{and}\quad \quad   P^-=P_{\infty}^- + P_0^-.
$$ 

\midinsert
\epsfxsize=250pt
\centerline{
\epsfbox{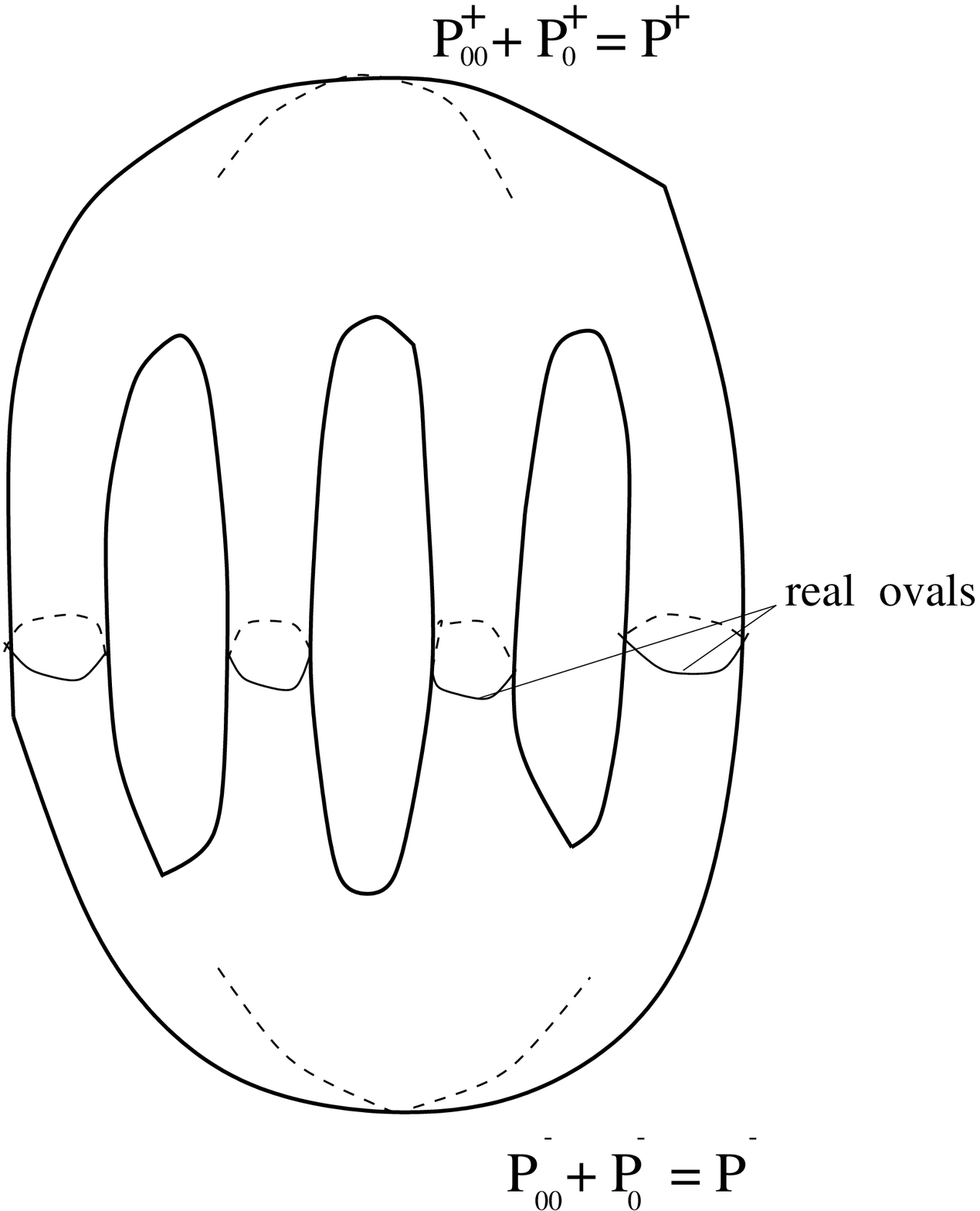}
}
\botcaption{fig. 4}
\endcaption
\endinsert

The cuts are the projections on the $\l$--plane of the real ovals of 
anti--holo\-morphic involution $\tau_a$. 
The infinities $P^+$ and $P^-$ are also fixed points of $\tau_a$. The 
multiplier $w(Q)$ is single-valued on $\Gamma$ and satisfies the identity 
$w(\tau_aQ) =\overline{w(Q)}$. This guaranties {\it continuity} 
of $w(Q)$ at the points $P^{+/-}$,  where $w(Q) \longrightarrow \infty/0$ 
respectively. The involution $\tau$ is defined so  that it preserves the 
infinities.  The formula 
$ w(Q)= \Delta(\l(Q))+ y(Q)$ implies $w(\tau Q)= (-1)^N w(Q)$. 

Let us introduce a multivalued function $p(Q)$ on $\Gamma$ by the formula $w(Q)=e^{p(Q)N}$. Obviously, 
$p(Q)$ is defined up to the integer multiple of $2\pi i/N$. The differential $dp$ is of the third kind with poles at 
$P^+$ and $P^-$. The relation $w(Q)w(\tau_{\pm}Q)=1$ implies that $p(Q)\equiv -p(\tau_{\pm}Q)$ (mod $2\pi i/N$). 
Therefore, in the vicinity of $P_{\infty}^{\pm}$,    
$$
\align
p_{\infty}^+(\l)& = + \log \l +I_0 + {I_1\over \l} + {I_2\over \l} + \cdots, \\
p_{\infty}^-(\l)& = -\log \l -I_0 - {I_1\over \l} - {I_2\over \l} + \cdots,
\endalign
$$
where $\l=\l(Q)$. The fact that $\tau$ preserves $P_{\infty}^{+/-}$ and 
$w(\tau Q) = (-1)^N w(Q)$ implies  $I_k=0$ 
for $k$ odd, so 
$$
\align
p_{\infty}^+(\l)& = + \log \l +I_0 + {I_2\over \l^2} + {I_4\over \l^4} + \cdots, \\
p_{\infty}^-(\l)& = -\log \l -I_0 - {I_2\over \l^2} - {I_4\over \l^4} + \cdots. 
\endalign
$$
The anti-involution $\tau_a$ maps $P_{\infty}^{+/-}$ to $P_{0}^{+/-}$.  This 
and relation $w(\tau_aQ)=\overline{w(Q)}$ imply, in 
the vicinity of $P_0^{\pm}$, 
$$
\align
p_{0}^+(\l)& = - \log \l +\overline{I}_0 + \overline{I}_2 \l^2 + \overline{I}_4 \l^4 + \cdots, \\
p_{0}^-(\l)& = +\log \l -\overline{I}_0 - \overline{I}_2 \l^2 - \overline{I}_4 \l^4 + \cdots. 
\endalign
$$
The explicit form of the integrals $I_0, I_2,\cdots$ will be computed in 
the next section.

\subhead 4. Integrals of motion\endsubhead 
To simplify the calculations, we introduce $V_{\text{new}}(n)=\sqrt{R_n}
V_{\text{old}}(n)$. The corresponding mulitipliers  are
$w_{\text{new}}=\sqrt{D} w_{\text{old}}$, and
$$
p_{\text{old}}=p_{\text{new}}-{1\over 2N} \log D.
$$
We write all formulas in this section for the {\it new} spectral problem.

We  introduce $T_{n,m}=
(1+W_n(\l))e^{Z_{n,m}(\l)}(1+W_m(\l))^{-1}$, where $Z$ is a diagonal matrix and 
$W$ is antidiagonal. Separating the diagonal and antidiagonal 
part in the equation $T_{n+1,m}= V(n) T_{n,m}$, we obtain
$$
\align
&e^{Z_{n+1,m}- Z_{n,m}}=\Lambda + \Psi_n W_n, \tag1 \\
&W_{n+1} e^{Z_{n+1,m}- Z_{n,m}} =\Lambda W_n+ \Psi_n, \tag2 
\endalign
$$
where
$$
\Lambda=\[\matrix   \l & 0\\
                     0 & \l^{-1} \endmatrix \] \quad \text{and} \quad 
\Psi_n=\[\matrix 0 & \psi_n\\
                \bpsi_n & 0\endmatrix \].
$$ 
Let $W_n(\l)= \sum\limits_{k=1}^{\infty} W_n^{k} \l^{-k}$, where 
$$
W_n^k=\[\matrix 0 & w_n^{k+}\\
                w_n^{k-} & 0 \endmatrix \].
$$
From  (1-2),  we have 
$
W_{n+1}(\Lambda + \Psi_n W_n) = \Lambda W_n + \Psi_n.
$
Introducing 
$$
\sigma_+=\[\matrix 1 & 0\\
                   0 & 0\endmatrix \] \quad \text{and} \quad 
\sigma_-=\[\matrix 0 & 0\\
                   0 & 1 \endmatrix \],
$$
we can write $\Lambda= \sigma_+ \l + \sigma_- \l^{-1}$ and 
$$
\aligned
\sum\limits_{k\geq 1} W_{n+1}^k \l^{-k} (\l\sigma_+ & + \l^{-1}\sigma_-) + 
\sum\limits_{k,p\geq 1} W_{n+1}^{r} \Psi_n W_n^p \l^{-r-p} \\
&= (\l \sigma_+ + \l^{-1}\sigma_-)\sum\limits_{k\geq 1} W_n^k \l^{-k}+\Psi_n. 
\endaligned \tag3
$$
Collecting terms with $\l^0$ in (3), we have 
$
W_{n+1}^1\sigma_+=\sigma_+W_n^1 +\Psi_n
$
and
$$
W_n^1=\[\matrix 0 & -\psi_n\\
                \bpsi_{n-1} & 0 \endmatrix \].
$$
Collecting terms with $\l^{-1}$ in (3), we have 
$ W_{n+1}^2 \sigma_+= \sigma_+ W_n^2$.
This implies \break $W_n^2\equiv 0$.  

Collecting terms in (3) of the order $\l^{-k},\; k\geq 2$, 
we have the reccurence formula
$$
W_{n+1}^{k+1} \sigma_+ - \sigma_+W_n^{k+1}= 
\sigma_- W_{n}^{k-1}- W_{n+1}^{k-1}\sigma_- - 
\sum\Sb r+p=k,\\ r,p\geq 1\endSb  W_{n+1}^{r} \Psi_n W_n^p; \tag4 
$$
whence, for $k=2$, 
$$
W_n^3=\[\matrix 0 & -\psi_{n+1}+ \psi_{n+1}|\psi_n|^2\\
                \bpsi_{n-2}-\bpsi_{n-2}|\psi_{n-1}|^2 & 0 \endmatrix \].
$$
Therefore, for $n=N$ and $m=0$ we have
$$
T_{N} (\l) =(1+W_{N}(\l)) e^{Z_{N,0}(\l)} (1+W_0(\l))^{-1}
$$
and 
$$
e^{Z_{N,0}}(\l)=\prod\limits_{r=0}^{N-1}(\Lambda+\Psi_r W_r).
$$
Using the fact that $\Psi_rW_r$ is diagonal we infer that  for 
$Q\in (P_{\infty}^+)$
$$
w(Q)=e^{p_{\infty}^+(\l)N} = \prod\limits_{r=0}^{N-1}\(\l + \psi_r 
\sum\limits_{k\geq 1} w_r^{k-}\l^{-k}\).
$$

Therefore,
$$
\align
p_{\infty}^+(\l)N& = N \log \l + \\
&+ {1\over \l^2} \sum\limits_{r=0}^{N-1} 
\psi_r w_r^{1-} + {1\over \l^4}\( \sum\limits_{r=0}^{N-1} \psi_r w_r^{3-} 
- {1\over 2}(\psi_r w^{1-}_r)^2\) + \cdots.
\endalign
$$
Using the explicit expression of  $W_n^1, W_n^3$, we have\footnote"*"{We 
write the coefficients for the {\it old} spectral problem.}
$$
\align
NI_0& \equiv  -{1\over 2} \log D ,  \\
NI_2&     =   \sum\limits_{r=0}^{N-1} \psi_r \bpsi_{r-1},\\
NI_4&     =\sum\limits_{r=0}^{N-1} \psi_r \bpsi_{r-2}\(1-|\psi_{r-1}|^2\) - 
{1\over 2} (\psi_r \bpsi_{r-1})^2.
\endalign
$$

\subhead 5. Floquet solution. Baker--Akhiezer function\endsubhead
The Floquet solution was defined in section 3 as a special solution of 
the spectral problem $\g_{n+1}=V(n) \g_n$ with the property 
$\g_N=T_N\g_0=w\g_0$. It is uniqlue specified by the boundary conditions 
$g_n^1 +g^2_n|_{n=0}=1$. 

\noindent
{\bf Example:} vanishing potential $\psi_n=0$. Then,
$$
V_2(n,\l)=\[\matrix \l & 0\\
                  0  & \l^{-1}\endmatrix \],  \quad \quad \quad \text{and} 
\quad \quad T_N(\l)= \[\matrix \l^{N} & 0\\
                  0  & \l^{-N}\endmatrix \].
$$
Therefore,\footnote"**"{For any $\g=\[\matrix a\\ c \endmatrix\]$ we  write 
$\gh=\[\matrix \bar{c}\\ \bar{a} \endmatrix\]$.} for $Q\in 
(P_{\infty}^{+}/P_0^{+}),$ 
$$
\g_n(Q) = e^{\pm n \log \l} \[ \g_0^+/\gh_0^+\], \quad\quad\quad w(Q)= e^{\pm N \log \l};
$$
and for $Q\in (P_{\infty}^{-}/P_{0}^-)$
$$
\g_n(Q) = e^{\mp n \log \l} \[ \g_0^-/\gh_0^-\], \quad\quad\quad 
w(Q)= e^{\mp N \log \l};
$$
vectors $\g_0^+,\; \g_0^-$ are 
$$
\g_0^+= \[\matrix 1\\0 \endmatrix \],\quad\quad\quad \g_0^- =\[\matrix 0\\1\endmatrix \].
$$

We  introduce another Floquet solution $\gt_n$ normalised by the condition 
$i\gt^1_n- i \gt_n^2|_{n=0}=1.$
Let us define
$$
D_0=1,\quad \quad \quad D_n=\prod\limits_{k=0}^{n-1} R_k,\quad n=1,\hdots, 
N-1; \quad \quad D_N=D.
$$

\proclaim{Lemma 1} i. The Floquet solution $\g_n(Q)$ is:
$$
\g_n(Q)=C(Q)\[\matrix T_n^{11}(\l)\\ T_n^{21}(\l) \endmatrix\] +(1-C(Q)) 
\[\matrix T_n^{12}(\l)\\ T_n^{22}(\l) \endmatrix\], \quad \quad \l=\l(Q).
$$
with
$$
C(Q)={T^{12}\over T^{12}- T^{11} +w(Q)}= {w(Q) -T^{22}\over T^{21}- 
T^{22} +w(Q)},
$$
here,  $T=T_N$.

ii. The Floquet solution $\gt_n(Q)$ is 
$$
\gt_n(Q)=R(Q)\[\matrix T_n^{11}(\l)\\ T_n^{21}(\l) \endmatrix\] +(i+ R(Q))
\[\matrix T_n^{12}(\l)\\ T_n^{22}(\l) \endmatrix\], \quad 
$$
with
$$
R(Q)={iT^{12}\over w(Q)- T^{11}- T^{12}} = { iT^{22} - iw(Q)\over w(Q)- T^{21}- T^{22}}.
$$

iii. The following formulas hold:
$$
\sigma_1 \g_n (\tau_a Q) =\bar{\g}_n(Q), \quad \quad \quad \sigma_1 \tilde{\g}_n(Q) 
(\tau_aQ)=\overline{\tilde{\g}_n(Q)}.
$$
and
$$
\g_n(\tau Q) = (-1)^n i \sigma_3 \tilde{\g}_n(Q).
$$

(iv). The Floquet solution $\g_n(Q)$ has $2N$ poles at the points $\gamma$'s on the real 
ovals of the curve $\Gamma$. On each oval there is just one $\gamma$.

(v). If $Q\in (P_{\infty}^+/P_0^+)$, then $\g_n(Q)$ has the development 
$$
\g_n(Q)=e^{\pm n\log \l}{1\over \sqrt{D_n}} \[ \sum\limits_{s=0}^{\infty} \g_s^+(n)\l^{-s} 
/ \sum\limits_{s=0}^\infty \hat{\g}^+_s (n) \l^s\],
$$
with 
$$
\g_0^+ =\[\matrix 1\\ 0\endmatrix\],\quad \quad \quad \g_1^+(n) = 
\[ \matrix -\psib_{-1}\\ \;\;\;  \psib_{n-1}\endmatrix \].
$$
Also for $Q\in (P_{\infty}^-/P_0^-)$, 
$$
\g_n(Q)= e^{\mp n \log \l} \sqrt{D_n} \[ \sum\limits_{s=0}^{\infty} \g_s^-(n)\l^{-s}
/ \sum\limits_{s=0}^\infty \hat{\g}^-_s (n) \l^s\],
$$
with 
$$
\g_0^- =\[\matrix 0\\ 1\endmatrix\],\quad \quad \quad \g_1^-(n) = 
\[ \matrix -\psi_{n}\\ \;\; \psi_{0}\endmatrix \].
$$
\endproclaim
\demo\nofrills{Proof.\usualspace} i. Obviously,
$$
\g_n(Q)= C'(Q)\[\matrix T_n^{11}(\l) \\ T_n^{11}(\l) i\endmatrix\] +C''(Q)
\[ \matrix T_n^{12}(\l) \\ T_{n}^{22}(\l) i\endmatrix \]
$$
with some $C'(Q)$ and $C''(Q)$. The normalisation $g_n^1 +g_n^2|_{n=0} =1$ 
implies $C''(Q)= 1-C'(Q)$. Another condition $T_n\g_0 =  w \g_0$ produces 
the system
$$
\[\matrix T_N^{11} & T_N^{12}\\
          T_N^{21} & T_N^{22}\endmatrix \] \[ \matrix C\\ 1-C \endmatrix \] = 
w \[ \matrix C\\ 1-C\endmatrix \].
$$
Solving the system, we obtain the stated formulas for $C(Q)$.

ii. The proof is identical to the proof of (i). 

iii. The proof is based on explicit formulas of (i) and (ii). Using (1) of 
section (3), we have
$$
T_{n,0}(\l (\tau_a Q))=\sigma_1 \overline{T_{n,0}(\l(Q))} \sigma_1 \quad 
\quad \quad \text{and} \quad \quad w(\tau_aQ)=\overline{w(Q)}.
$$
From this we have $1-\overline{C(\tau_aQ)}=C(Q)$.  The proof is finished by 
substituting $T_n(\l(\tau_aQ))$ and $C(\tau_aQ)$ into explicit formula of (i). 
Similarly  $\overline{R(\tau_aQ)}=i +R(Q)$.  This implies the stated formula 
for $\tilde{\g}_n$. 

The last formula for $\g_n(\tau Q)$ is proved along the same lines using 
(2) and (6) of section 3. 

(iv). It is easy to see, by perturbation arguments, that $\mu_k=\l(\gamma_k)$ 
are near the points $\l^{\pm}_k= e^{i{2\pi\over 2N}k},\; k=0,\hdots, 2N-1$ 
for a small potential.  We will prove that $|\mu_k| =1$ {\it always}. 

Let $\g_n(Q)$ have a pole at some point $\gamma$. Then, from the formula of 
(i),  we have
$$
T^{11}- T^{12} = w(Q) \quad \quad \quad \text{and} 
\quad \quad  T^{22}- T^{11}=w(Q).
$$
Consider the special solution $\f_n(\mu)=T_n^{(1)}(\mu) - T_n^{(2)}(\mu)$ 
of the eigenvalue problem $\Lambda_n \f_n= \mu \f_n$. We have
$$
\f_0=\[\matrix 1\\ -1\endmatrix \] \quad \quad \quad \text{and} \quad \quad 
\f_N =\[\matrix w(Q)\\ -w(Q) \endmatrix \].
$$
Now the Cauchy formula (7) of section 3 to obtain 
$$
<\Lambda \f, \f>_{\heartsuit} = <\f,\Lambda^{-1} \f>_{\heartsuit}+
\text{boundary terms}.
$$
Note, that $\Lambda\f =\mu \f, \;\; \Lambda^{-1} \f =\mu^{-1} \f$.  
It is easy to compute the boundary terms: to wit,  
$$
{1\over N} (|w|^2-1) (\bar{\mu}^{-1}- \mu).
$$
Substituting this into the Cauchy formula,  we obtain
$$
(1-|\mu |^2) \[ {1\over N} (|w|^2 -1) - (\f,\f)\]=0.
$$
This implies $|\mu|=1$.

(v). Consider $P_{\infty}^+/ P_0^+$. The relation between two asymptotic 
expansions follows from the formula $\sigma_1 \g_n(\tau_a Q)=\g_n (Q)$. 
The actual form of the coefficient is derived by substituting the asymptotic 
expansion $$\g_n(Q)= e^{n\log \l} {1\over \sqrt{D_n}} \sum\limits_{s=0}^{\infty} \g_s^+(n) \l^{-s}$$ into the spectral problem $\g_{n+1} = 
V(n,\l)\g_n$, where $V(n,\l) =[ \l \sigma_+ +\l^{-1} \sigma_- $ $ +\Psi_n].$ 
We arrive at  the reccurence relation
$$
\sigma_+ \g_{s+1}^+(n) + \sigma_- \g_{s-1}^+ (n) +\Psi_n \g_s^+(n) =\g_{s+1}^+(n+1).
$$
The boundary condition $g_n^1 +g_n^2 |_{n=0}=1$ implies 
$$
g_0^1(n) + g_0^2(n)|_{n=0} =1; \quad \quad \quad \quad \quad \quad 
g_s^1(n) + g_s^2(n)|_{n=0}=0 \quad \quad \text{for} \quad \quad  s\geq 1.
$$ 
Starting from $s=-1$, reccurently, one can compute $\g_0^+(n), \; \g_1^+(n), \, etc.$ 
\qed
\enddemo

On the curve $\Gamma$ one can consider  the Baker-Akhiezer (BA) function 
$\g(\tau,n,t,\l)$ with $2N$ poles on the real ovals. On each oval there 
is just one pole. The function 
$\g(\tau,n,t,\l)$ has the following assymptotics at the infinities:
$$
\alignat 2
\g(\tau,n,t,\l)&= e^{{i\over 2}\tau  +n\log \l + i(\l^2 -1) t} {1\over \sqrt{D_n}}
\sum\limits_{s=0}^{\infty} \g_s^+ \l^{-s},\quad \quad \quad &Q\in (P_{\infty}^+),\\
\g(\tau,n,t,\l)&= e^{-{i\over 2}\tau  -n\log \l - i(\l^{-2} -1) t} \sqrt{D_n} 
\sum\limits_{s=0}^{\infty} \g_s^- \l^{-s},\quad \quad \quad &Q\in (P_{\infty}^-),\\
\g(\tau,n,t,\l)&= e^{-{i\over 2} - n\log \l - i(\l^{-2}  -1) t} {1\over \sqrt{D_n}} 
\sum\limits_{s=0}^{\infty} \hat{\g} _s^+ \l^{s},\quad \quad \quad &Q\in (P_{0}^+),\\
\g(\tau,n,t,\l)& = e^{{i\over 2} +n\log \l + i(\l^2 -1) t}  \sqrt{D_n} 
\sum\limits_{s=0}^{\infty} \hat{\g} _s^- \l^{s},\quad \quad \quad &Q\in (P_{0}^-).
\endalignat
$$
The BA function with these properties exists and 
defined uniqly. 
The BA function $\g(\tau, n, t, Q)$ satisfies the identities
$$
\align
&[\partial_{\tau} -V_1] \g(\tau, n, t, Q)= 0,\\
&[\Delta  -V_2] \g(\tau, n, t, Q)= 0,\\
&[\partial_{t} -V_3] \g(\tau, n, t, Q)= 0.
\endalign
$$
The explicit form of the $V$'s is given in section 2. The Floquet solution 
is a particular case of BA function with the variables $\tau, t$ fixed.  

\subhead 6. Dual Floquet solution. Variational identity\endsubhead 
One can write the equation $V(n)\g_n =\g_{n+1}$ in the form\footnote"*"{$J=i\sigma_2$.} 
$$
[J\Delta- J V(n)]\g_n=0.
$$
Let us define the dual Floquet solution $\g^+=\[g^{1+}, g^{2+}\]$ at the 
point $Q$  by 
$$
\g^+_n(Q)=\g_n(\tau_{\pm}Q)^T.
$$
\proclaim{Lemma 2} The dual Floquet solution $\g_n^+(Q)$ satisfies\footnote"**"{$\f_n 
\Delta= \f_{n-1}.$}
$$
\g^+_n(Q)\[J\Delta-JV(n-1,\l)\]=0.
$$
\endproclaim
\demo\nofrills{Proof.\usualspace} The equation $V(n-1)\g_{n-1}=\g_{n}$ can 
be written as $\g_{n-1}= V^{-1}(n-1)\g_n$, where
$$
V^{-1}(n-1)={1\over \sqrt{R_{n-1}}}\[\matrix \l^{-1} & - \psi_{n-1}\\
                                            -\psib_{n-1} & \l \endmatrix \];
$$
or, in coordinates,
$$
\align
g_{n-1}^1&={1\over \sqrt{R_{n-1}}}(\l^{-1} g_n^1 - \psi_{n-1} g_n^2),\\
g_{n-1}^2&={1\over \sqrt{R_{n-1}}}(-\psib_{n-1} g_n^1 +\l g_n^2).
\endalign
$$
We can also rewrite it in the form
$$
(g_n^1,g_n^2)\(\[\matrix 0 & \Delta \\
                  -\Delta & 0  \endmatrix\] - {1\over \sqrt{R_{n-1}}} 
\[\matrix \psib_{n-1} & \l^{-1} \\
  -\l & -\psi_{n-1} \endmatrix \] \) =0.
$$
This is exactly the stated identity.
\qed
\enddemo

We need the standard formula \cite{KP} for variations of the quasi-momentum.

\proclaim{Lemma 3} i. The expression $\g^+_n(Q)J\g_n(Q)$ does not depend on n and
$$
\g^+_n(Q)J\g_n(Q)=< \g^+J \g> =\Psi(Q).
$$

ii. The following  identity holds
$$
\d p \Psi (Q)= <\g^+_n J \d V(n-1) \g_{n-1}>.
$$
\endproclaim
\demo\nofrills{Proof.\usualspace} i. Can be checked using difference equation 
for 
$\g_n(Q)$ and $\g_n^+(Q)$. 

ii. Denote by $\tilde{V}(n,\l)$ and $\gt_n(\l)$ deformed matrix $V(n,\l)$ and 
the Floquet solution $\g_n(\l)$; to wit,
$
\tilde{V}=V +\epsilon \delta V +o(\epsilon)$ and  $\gt=\g + \epsilon \delta 
\g + o(\epsilon)$.  Then,
$$
\g_n^+(Q)\[(J\Delta -J\tilde{V}(n-1,\l))\gt_{n-1}(Q)\]=0,
$$
$$
\[\g_n^+(Q)(J\Delta -JV(n-1,\l))\]\gt_{n-1}(Q)=0.
$$
Subtracting, we obtain
$$
\align
\sum\limits_{n=0}^{N-1}& \g_n^+(J\Delta \gt_{n-1}) -(\g_n^+ J\Delta) \gt_{n-1}\\
&=\sum\limits_{n=0}^{N-1}\g_n^+\(J\tilde{V}(n-1) \gt_{n-1} \)-\(\g_n^+ J V(n-1)\)\gt_{n-1}.
\endalign
$$
It is easy to see, that
$$
\text{RHS}= \epsilon \sum\limits_{n=0}^{N-1} \g_n^+ J \d V(n-1) \g_{n-1} + 
o(\epsilon).
$$
Using the formula
$$
\sum\limits_{n=0}^{N-1} (\g_n^+ J\Delta \g) \gt_{n-1}= \sum\limits_{n=0}^{N-1} 
\g_n^+ (J\Delta\gt_{n-1}) + \g_{-1}^+ J \gt_{-1} - \g_{N-1}^+ J \gt_{N-1},
$$
we obtain
$$
\text{LHS}=\g_{N-1}^+ J\gt_{N-1} - \g_{-1}^+J \gt_{-1}.
$$
From the definition of the Floquet solution 
$$
\g_{N-1}^+= e^{-Np}\g_{-1}^+\quad \text{and}\quad 
\gt_{N-1}=e^{N\tilde{p}}\gt_{-1},
$$
we have 
$$
\text{LHS}= \(e^{N(\tilde{p}- p)}-1\) \g_{-1}^+J \gt_{-1}=\epsilon N \d p \g_{-1}^+ 
J \g_{-1} + o(\epsilon).
$$
Collecting terems with $\epsilon$,  we obtain the stated identity
\qed
\enddemo

Let us introduce $\g^*_n(Q)$, the dual Floquet solution normalized by the 
condition $<\g^*J\g>=1$. Obviously,
$$
\g_n^*(Q)={\g^+_n(Q)\over \Psi(Q)};
$$
$\g_n^*(Q)$ has poles at the branch points of the curve.

\subhead 7. Hamiltonian formalism for the Ablowitz-Ladik system\endsubhead
As  shown in \cite{KP}, the formula
$$
\omega_0={i\over 2} 
\sum \res {1\over \Psi(Q)}<\g^+(n)J\delta V(n-1)\wedge \delta\g(n-1)> {d \l\over 
\l}.
$$
defines a closed, nondegenerate 2 form on the space of operators $\Delta- V_2$ with 
periodic potential. Our goal to compute the residue {\it explicitly}.

Near $P_0^+$,  
$$
{1\over \Psi(Q)}= \psi_0 + \psi_1 \l^1 + \cdots, 
\quad\quad\quad \quad \quad  \; Q\in (P_0^+). \tag1
$$
The identity
$$
\align
\Psi(\tau_{\pm}Q)=<\g^+(\tau_{\pm} Q) J \g(\tau_{\pm} Q)> & = 
<\g^T( Q) J \g^+( Q)^T>\\
& =- <\g^{+}( Q) J \g(Q)>= -\Psi(Q)
\endalign
$$
implies
$$
{1\over \Psi(Q)}= -\psi_0 - \psi_1 \l^1  - \cdots, \quad\quad\quad\quad\quad 
\; Q\in (P_0^-). \tag2
$$
The involution $\tau_a$ maps $P^{+/-}_{\infty}$ to $P^{+/-}_0$, 
and $\g_n(\tau_a Q)=\sigma_1 \bar{\g}_n(Q)$. Therefore, using $\tau_{\pm} 
\tau_{a}= \tau_{a}\tau_{\pm}$, we have
$$\align
\Psi(\tau_aQ)&=\g^T(\tau_{\pm}\tau_a Q) J \g (\tau_a Q)= 
\g^T(\tau_a\tau_{\pm} Q) J \g (\tau_a Q)\\
& =\bar{\g}^T(\tau_{\pm} Q) \sigma^T_1 J\sigma_1\bar{\g} (Q)= 
-\overline{\Psi(Q)}
\endalign
$$
so that, if $Q\in (P_{\infty}^+)$, then $\tau_a Q \in (P_0^+)$ and 
$$
\aligned
{1\over \Psi(Q)}&=-\overline{{1\over \Psi(\tau_a Q)}}
=-\overline{\(\psi_0 + 
{\psi_1 \over \bar{\l}^1} + \cdots\) }\\ 
&=-\psib_0 -{\psib_1\over \l^1}- \cdots, 
\quad\quad\quad \quad \quad \quad  \; Q\in (P_{\infty}^+). \endaligned \tag3
$$
Again, applying $\tau_{\pm}$, we have 
$$
{1\over \Psi(Q)}=\psib_0 +{\psib_1\over \l^1}+ \cdots,
\quad\quad\quad \quad \quad \quad  \; Q\in (P_{\infty}^-).  \tag4
$$
It is easy to compute
$$
\align
\psi_0&=<\gh^{-T}_0 J \gh_0^+>=1\\
\psi_1&=<\gh_0^{-T}J\gh_1^{+}> + <\gh_1^{-T}J\gh_0^{+}>= \psib_0-\psi_{-1}, \quad etc.
\endalign
$$

Similarly, introducing $S(Q)\equiv <\g_n^+(Q)J\d V(n-1)\wedge \d \g_{n-1}(Q)>$, we obtain
$$
S(\tau_a Q)=-\overline{S(Q)}. \tag5
$$

One can show that, in the vicinity of the infinities:
$$
\align
S(Q)&=c_0+c_1\l +\cdots, \quad \quad \quad Q\in(P_0^+),\\
S(Q)&=d_0+{d_{-1}\over \l} +\cdots, \quad \quad \quad Q\in (P_{\infty}^{+});
\endalign
$$
and

$$
\align
S(Q)&=a_0+a_1\l+\cdots \quad \quad \quad Q\in(P_0^-),\\
S(Q)&=b_0+{b_{-1}\over \l}+\cdots \quad \quad \quad Q\in(P_{\infty}^-).
\endalign
$$
(5) implies $d_{-k}=-\bar{c}_k$ and $b_{-k}=-\bar{a}_k$. 
Computing residues,  we have 
$$
\align
\underset{P_{\infty}^+}\to\res {S\over \Psi} {d\l\over \l} &= 
\underset{P_{\infty}^+}\to\res \(d_0+\cdots\) \(-\psib_0 -\cdots\) 
{d\l\over \l}= d_0 \psib_0,\\
\underset{P_{\infty}^-}\to\res {S\over \Psi} {d\l\over \l} &= 
\underset{P_{\infty}^-}\to\res \(b_0+\cdots\) \(\psib_0 +\cdots\) 
{d\l\over \l}=-b_0 \psib_0,
\endalign
$$
and
$$
\sum\limits_{P_{\infty}^{+/-}} \res {S\over \Psi} {d\l\over \l} =d_0-b_0.
$$
Similarly, 
$$
\align
\underset{P_{0}^+}\to\res {S\over \Psi} {d\l\over \l} &=
\underset{P_{0}^+}\to\res \(c_0+\cdots\) \(\psi_0 +\cdots\)
{d\l\over \l}= c_0 \psi_0,\\
\underset{P_{0}^-}\to\res {S\over \Psi} {d\l\over \l} &=
\underset{P_{0}^-}\to\res \(a_0+\cdots\) \(-\psi_0 -\cdots\)
{d\l\over \l}=-a_0 \psi_0,
\endalign
$$
and
$$
\sum\limits_{P_{0}^{+/-}} \res {S\over \Psi} {d\l\over \l} =c_0-a_0=-\overline{(d_0-b_0)}.
$$

Using the formulas for Floquet solutions from Lemma 1, we have
$$
\align
b_0 =& \;\;\; <{\g_0^{+T}\over \sqrt{D_n}} J \d {\sigma_-\over 
\sqrt{R_{n-1}}}\wedge \d 
\sqrt{D_{n-1}}\g_0^-> \\
 &+  <{\g_0^{+T}\over \sqrt{D_n}} J \d {\Psi_{n-1}\over 
\sqrt{R_{n-1}}}\wedge \d \sqrt{D_{n-1}}\g_1^->\\
 &+  <{\g_1^{+T}\over \sqrt{D_n}} J \d {\sigma_+\over \sqrt{R_{n-1}}}\wedge 
\d \sqrt{D_{n-1}}\g_1^->\\
 & +  <{\g_1^{+T}\over \sqrt{D_n}} J \d {\Psi_{n-1}\over \sqrt{R_{n-1}}}\wedge 
\d \sqrt{D_{n-1}}\g_0^->\\
=& \;\;\;  <{1\over \sqrt{D_n}} \d {1\over \sqrt{R_{n-1}}}\wedge \d \sqrt{D_{n-1}}>\\
& - <{1\over \sqrt{D_n}} \d {\psib_{n-1}\over \sqrt{R_{n-1}}}\wedge \d \sqrt{D_{n-1}} 
\psi_{n-1}>\\
& + <{\psib_{n-1}\over \sqrt{D_n}} \d {1\over \sqrt{R_{n-1}}}\wedge \d \sqrt{D_{n-1}}
\psi_{n-1}>\\
& - <{\psib_{n-1} \over \sqrt{D_n}} \d {\Psi_{n-1}\over \sqrt{R_{n-1}}}\wedge \d \sqrt{D_{n-1}}>.
\endalign
$$
After simple algebra,
$$
\therefore
= {1\over 4} <{\d R_{n-1}\over R_{n-1}} \wedge {\d D_{n-1}\over D_{n-1}}> - 
<{1\over R_{n-1}} \d \psib_{n-1}\wedge \d \psi_{n-1}>.
$$
Similarly,
$$
\align
d_0 &= < \sqrt{D_n} \g^{-T}_0 J \d {\sigma_+\over \sqrt{R_{n-1}}}\wedge \d {\g_0^+\over 
\sqrt{D_{n-1}}}>\\
&= -{1\over 4} <{\d R_{n-1}\over R_{n-1}} \wedge {\d D_{n-1}\over D_{n-1}}>.
\endalign
$$
Finally,
$$
\align
\sum\limits_{P_{\infty}^{+/-}} \res {S\over \Psi} {d\l\over \l} &= 
\;\; \; <{1\over R_{n-1}} \d \psib_{n-1} \wedge \d \psi_{n-1}> - 
{1\over 2}<{\d R_{n-1}\over R_{n-1}}\wedge {\d D_{n-1}\over D_{n-1}}>\\
\sum\limits_{P_{0}^{+/-}} \res {S\over \Psi} {d\l\over \l} &= 
- <{1\over R_{n-1}} \d \psi_{n-1} \wedge \d \psib_{n-1}> + {1\over 2} <{\d R_{n-1}\over 
R_{n-1}}\wedge {\d D_{n-1}\over D_{n-1}}>.
\endalign
$$
Taking the sum, we obtain
$$
\omega_0=<{i\over R_{n-1}} \d \psib_{n-1}\wedge \d \psi_{n-1}>.
$$

\proclaim{Lemma 5} (i) The formula
$$
\xi_n(Q)=\l^n<\g^* J\d V \wedge \d \g> {d \l\over \l}
$$
$n=\hdots,-1,0,1,\hdots,$ defines meromorphic in $Q$ differential form on $\Gamma$ with poles
at $\gamma_1,\cdots, \gamma_{2N}$ and $P_{\infty}^{+/-}, P_{0}^{+/-}$.

(ii) The symplectic 2-forms defined by the formula
$$
\omega_n= {i\over 2}\sum\limits_{P} \res \xi_n(Q)
$$
can be written as
$$
\omega_n= -{i\over 2} \sum\limits_{k=1}^{2N} \l^n(\gamma_k)  
\d p(\gamma_k) \wedge {\d \l\over \l}  (\gamma_k).
$$
\endproclaim
\demo\nofrills{Proof.\usualspace} (i) The poles of $\g_n^{*}$ at the branch points 
$(\l^{\pm},0)$ are killed by the zeros of $d\l$. The rest are just 
$\gamma_1,\cdots, \gamma_{2N}$ and $P_{\infty}^{+/-}$ and $P_{0}^{+/-}$. 

(ii) By Cauchy's theorem, 
$$
\sum\limits_{P} \res \xi_n(Q) + \sum\limits_{k=1}^{2N} \res \xi_n(Q)=0.
$$
Near $\gamma_k$,
$$
\g_n={\res \g_n\over \l-\l(\gamma_k)} +O(1).
$$
Therefore,
$$
\d \g_n(Q)= {\res \g_n\over (\l-\l(\gamma_k))^2} \d\l(\gamma_k) +O(1)=
{\g_n\over \l-\l(\gamma_k)} \d\l(\gamma_k) +O(1).
$$
Note that $\g^*_n(\gamma_k)=0$ and,  using Lemma 3,  
$$
\underset{\gamma_k}\to\res \xi_n(Q)=\l^n<\g^*J\d V \g>\wedge {\d\l\over \l}
(\gamma_k) \underset{\gamma_k}\to\res {d\l\over \l-\l(\gamma_k)}= 
\l^n(\gamma_k) \d p(\gamma_k) \wedge {\d \l\over \l} (\gamma_k).
$$
We are done.
\qed
\enddemo

The bracket $\{\bullet,\bullet\}_{\omega_0}$ is constructed from the symplectic 
form $\omega_0$
$$
\{f,g\}_{\omega_0}\equiv i\sum\limits_{n=1}^{N} R_n\( 
{\d f\over \d \psi_n} {\d g\over \d \psib_n}-  
{\d f\over \d \psib_n} {\d g\over \d \psi_n}\).
$$
The original AL flow from section 2 can be written 
$$
\overset\bullet\to\psi_n=\{ \psi_n, H\}_{\omega_0}, \quad \quad \quad H
%=\sum \psi_n \psib_{n+1} + \psi_{n} \psib_{n-1} + 2 R_n 
= N\( I_2 + \overline{I}_2 - 2 I_0 - 2\overline{I}_0\).
$$
The phase flow is also Hamiltonian
$$
\overset\bullet\to\psi_n=\{ \psi_n, P\}_{\omega_0}, \quad \quad \quad P=
N(I_0+ \overline{I}_0). 
$$

\subhead 8. Embedding of the AL system into the function space\endsubhead
First, we introduce   interpolating trigonometrical polynomials.  

Let $f(x)$ be a smooth 1-periodic complex function and  
associate to it  the sequence of interpolating trigonometrical  polynomials  
$$
f_N(x)=\sum_{|k|\leq m} e^{2\pi i k} \hat{f}_N(k),\quad \quad \quad  
N=2m+1,\;\; m=1,2\hdots.
$$ 
These have the property that $f_N(x)= f({x})$ for 
$x\in \T_N=\{x\in\T: x={n\over N},\; n=0,\hdots, N-1\}$. Let $\M_N$ 
be a space of  such trigonometric polynomials of degree $m$; it is in  
one-to-one correspondence with the space $M_N$ of 
N-periodic complex sequences: 
$$
f_N (x) \longleftrightarrow f_n, \quad  
f_N (x) \in \M_N, f_n \in M_N\quad  $$
if we put $ \epsilon f_N({n\over N})=f_n, \;\; n=0,\hdots, N-1.$

Introduce $\H_1^N(\psi)\equiv H_1^N(\epsilon \psi)$. The region    
$$
\B^N=\{ \psi_N\in \M_N:\; N\H_1^N(t \psi)<\infty, \;\;\text{for all} \;\; 
0\leq t\leq 1\}
$$
is called the "box" of the space $\M_N$; the trigonometric 
polynomial $\psi_N$ belongs $\B^N$ if and only if $|\psi_N(x)|<N, \;x \in \T_N  $.  
The map $\psi_N(x) \leftrightarrow \psi_n$ allows us to define the  
flow $e^{t\X_3^N}$ on  $\B^N$ by the formula  
$$
e^{t\X_3^N}\psi_N\equiv \epsilon^{-1}e^{\theta X_3^N} \epsilon \psi_N,\quad \quad 
\quad \theta=\epsilon^{-2}t.
$$
Evidently  $\B^N$ is invariant under the flow $^{t\X_3^N}$. Due to the 
natural embedding $\M_N\subset \M$,  the dynamics $e^{t\X_3^N}$ also can be 
defined by the formula $e^{t\X_3^N}\psi\equiv e^{t\X_3^N}\psi_N$ for any 
function $\psi\in \M$ which satisfies the inequality 
$|\psi(x)|< N,\; x\in \T_N$. 

We  define  on $\M_N$ the volume form 
$$
d\,vol^N\equiv  {1\over N! \D_N(\psi)} \underset{x\in \T_N}\to\bigwedge 
i\d\psi(x) \wedge \d \psib(x),
$$
where $\D_N(\psi)\equiv D_N(\epsilon \psi)$ and the functional 
$\H_5^N(\psi)$ is by definition $H_5^N(\epsilon\psi)$. The flow 
$e^{\theta X_3}$ preserves both $H_3(\psi_n,\psib_n)$ and the volume form
$$
d\, \text{vol}= {1\over N! D_N}\, \underset{n}\to\bigwedge\,i 
\d\psi_n\wedge\d \psib_n.
$$
\newpage

\noindent
Therefore, the volume form $d\,vol^N= \epsilon^{- 2N}d\,\text{vol}$ and the 
functional 
$\H_5^N$ are  invariant under the flow $e^{t\X_3^N}$.
The  finite measure $d \mu^N(\psi,\psib)$ on $\B^N$  with density
$$
 e^{-{N^5\over 2} \H_5^N(\psi)}d\, vol^N=
e^{-{N^5\over 2}\H_5^N(\psi)} {1\over N! \D_N(\psi)} 
\underset{x\in \T_N}\to\bigwedge i\d \psi(x)\wedge \d \psib(x)
$$
is also invariant under the flow $e^{t\X_3^N}$.

To ensure proper analytic control\footnote"*"{Similar cut-off was introduced in \cite{MCV2}.} we need to introduce  the function $h(x)$ (fig. 5) 
\midinsert \epsfxsize=250pt
\centerline{
\epsfbox{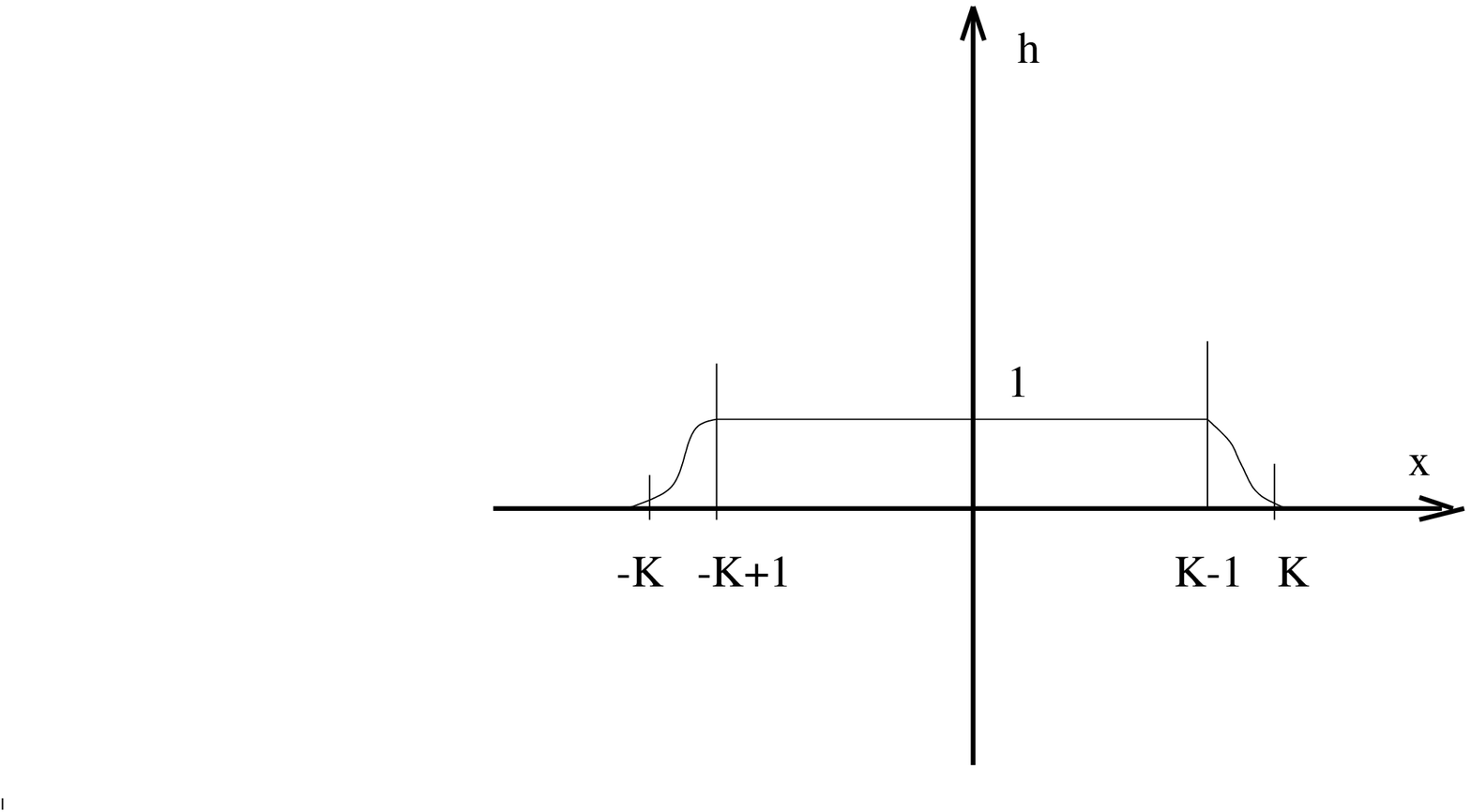}
}
\botcaption{fig. 5}
\endcaption
\endinsert
\noindent
and we define the invariant functional $\chi_K^N(\psi)\equiv h(N\H_1^N) 
h(-N^3\H_3^N)$.  On the box $\B^N$ the probability measure  
$d\mu^N_K(\psi,\psib)$ 
$$
{1\over \Xi_K^N} \chi_K^N(\psi) e^{-{N^5\over 2} \H_5^N(\psi)} d\, vol^N,
$$
where $\Xi_K^N$ is a normalisating factor\footnote"**"{The 
precise definition of $\Xi_K^N$ and $\Xi_K$ will be given in Lemma 8.}.

Due to the  embedding $\M_N\subset \M$,  the measure $d\mu^N_K(\psi,\psib)$ 
can be transfered to the whole of $\M$. We  define on $\M$ the probability 
measure $d\mu_K(\psi,\psib)$ with the density 
$$
{1\over \Xi_K} \chi_K(\psi) e^{-{1\over 2} \H_5}d\, vol=
{1\over \Xi_K} \chi_K(\psi) e^{-{1\over 2} \H_5} {1\over \infty!}
\, \underset{x\in \T}\to\bigwedge\, i \d \psi(x) \wedge \d \psib(x),
$$
where $\chi_K(\psi)=h(\H_1)h(\H_3)$ and $\Xi_K$ is a normalisating factor.   

In order to prove  invariance of the measure $d\mu_K$, we use the method of 
weak solutions introduced by McKean, \cite{MC}. 
The measure $d\mu_K^N$ on the initial data 
defines the measure $d \m^N_K$ in the space of paths 
$C\([0,T]\rightarrow \M\)$ for  $T>0$. 
All information about the measure $d\mu_K^N$ and the flow $e^{t\X_3^N}$ 
is encoded now into the measure $d\m_K^N$.  The proof of invariance takes  
three steps.

\noindent
Step 1. The measures  $d\mu^N_K$ converge to 
$d \mu_K$  weakly in $H^s,\; 
1\leq s<{3\over 2}$, as $N\rightarrow \infty$.  

\noindent
Step 2. For any fixed $K$,  the family  $d\m_K^N,\; N=1,2\hdots$ 
is tight and converges to  measure $d\m_K$. The stationary measure 
$d\m_K$  for fixed $t$  has marginal distribution $d\mu_K$. 

\noindent
Step 3. The measure  $d\m_K$ is supported on the solutions 
of the NSL flow. 

All three steps will be completed in subsequent sections. 
From the invariance of the measure $d\mu_K$, it is easy to infer the   
invariance of the the desired Gibbs' state with the density  
${1\over \Xi} e^{-{1\over 2}\H_5} d\, vol$.

\subhead 9. Convergence of AL Gibbs' state to the NLS Gibbs' state\endsubhead
In the previous section,  we introduced the family of probability measures 
$d\mu^N_K(\psi,\psib)$ on $\B^N\subset \M$ with the 
densities\footnote"*"{To simplify notation, we omit the normalization 
factors $\Xi$ for a moment.}
$$
\chi_K^N(\psi) e^{-{N^5\over 2} \H_5^N(\psi)}d\,  vol^N = \chi_K^N(\psi)
e^{-{N^5\over 2} \H_5^N (\psi)} {1\over N!\,  \D_N(\psi)} \underset{x\in 
\T_N}\to\bigwedge i\d\psi(x) \wedge \d \psib(x).
$$
We will show that $d\mu^N_K \rightarrow d\mu_K,\,\text{weakly in } \, H^s,\; 
1\leq s<{3\over 2}$,  as $N\rightarrow \infty$, where $d\mu_K(\psi,\psib)$ 
has the density
$$
\chi_K(\psi) e^{-{1\over 2}\H_5(\psi)}d\, vol= \chi_K(\psi) 
e^{-{1\over 2}\H_5(\psi)} 
{1\over \infty!} \underset{x\in \T}\to\bigwedge i \d\psi(x)\wedge \d\psib(x).
$$
The cut-off in $K$ can be removed easily and $d\mu_K $ converges to  $d\mu$.

We  split the integral $H_5^N$ into a quadratic part $G^N(\psi)$ and  a
nonlinear part $R^N(\psi)$.  Using the explicit expression for the integrals 
$I$'s,  we have 
$$
\align
H_5^N& (\psi_n)= G^N(\psi_n) + R^N(\psi_n)\\ 
&       =\,\;  \sum(\psi_{n+1} \psib_{n-1} + \psib_{n+1} \psi_{n-1}) 
- 4(\psi_n \psib_{n+1} + \psib_n \psi_{n+1}) + 6|\psi_n|^2\\
       &\;\;  -\sum(\psi_{n+1}\psib_{n-1} + \psib_{n+1} \psi_{n-1}) |\psi_n|^2 
       -{1\over 2} \sum (\psi_n\psib_{n-1})^2 + (\psib_n \psi_{n-1})^2 \\
&\;\; - \sum 6|\psi_n|^2 + 12 N I_0.
\endalign
$$ 
It is easy to check that 
$$
G^N(\psi_n)=\sum\limits_{n=0}^{N-1} |\psi_{n+1} -2 \psi_n + \psi_{n-1}|^2. 
$$
Introduce
$$
I^N(\psi_n)= \sum\limits_{n=0}^{N-1}|\psi_n|^2
$$
and also
$$
\G^N(\psi)=G^N(\psi_n),\quad \R^N(\psi)=R^N(\psi_n),
\quad \I^N(\psi)=I^N(\psi_n),
$$
where $\psi_n=\epsilon \psi({n\over N})$. 
With the new notations, 
$$
\align
\chi_K^N(\psi) e^{-{N^5\over 2}\H^N_5(\psi)}  d\, & vol^N\\ 
 =& \chi_K^N(\psi){1\over\D_N(\psi)}e^{-{1\over 2}
\[-N\I^N(\psi)+N^5\R^N(\psi)\]}\\
&\times  e^{-{1\over 2} \[ N^5 \G^N(\psi) + N\I^N(\psi)\]}\,  {1\over N!} 
\underset{x \in \T_N}\to\bigwedge i \d\psi(x)\wedge \d \psib (x)\\ 
\asymp & \chi_K^N(\psi){1\over \D_N(\psi)}e^{-{1\over 2} \W^N(\psi)}\times 
d\gamma^N (\psi,\psib).
\endalign
$$
Now we are ready to prove 
\proclaim{Lemma 6} The family of Gaussian probability measures 
$d\,\gamma^N(\psi,\psib)$ with the densities 
$$
e^{-{1\over 2} \[N^5\G^N(\psi)+ N\I^N(\psi)\] }{1\over N!}\underset{x \in 
\T_N}\to\bigwedge i \d\psi(x)\wedge \d\psib(x)
$$
converges weakly in $H^s,\, s<{3\over 2}$, as $N\rightarrow \infty$,  to the 
probability measure $d\gamma(\psi,\psib)$: 
$$
e^{-{1\over 2} \int |\psi''|^2 +|\psi|^2} {1\over \infty!} \underset{x \in \T}\to
\bigwedge i \d\psi(x)\wedge\d\psib(x).
$$
\endproclaim  
\demo\nofrills{Proof.\usualspace}
For a smooth $\psi(x) =\sum\limits_{k} e^{2\pi i x k} 
\hat{\psi}(k)$ and $N\rightarrow \infty$, 
$$
\align
N^5  & \G^N  (\psi)   +N\I^N(\psi)\\ 
 & = N^5 \epsilon^2\sum |\psi\(n+1/N\) - 
2\psi\(n/ N\) + \psi\(n-1/ N\)|^2 +N\epsilon^2\sum 
|\psi\(n/N\)|^2\\
& \approx N^3 \epsilon^4 \sum |\psi''\(n/ N\)|^2+\epsilon\sum |\psi\(n/ N\)|^2
\rightarrow  \int\limits_{0}^{1} |\psi''(x)|^2 + |\psi(x)|^2 d\,x.
\endalign
$$ 
This explains on a formal level, the convergence $d\, \gamma^N\rightarrow d\, 
\gamma$.
 
To obtain the full proof, we write all measures in terms of Fourier 
coefficients.  First we note that
$$
\int\limits_{0}^{1}|\psi''(x)|^2 +|\psi(x)|^2 dx=
\sum\limits_{k} (16\pi^4 k^4 +1) |\hat{\psi}(k)|^2=\sum\limits_{k}
\sigma^2(k)|\hat{\psi}(k)|^2.
$$
Using the identity 
$$
{1\over N}\sum\limits_{p=0}^{N-1} |\psi(n/N)|^2= 
\sum\limits_{|k|\leq m} |\hat{\psi}_N(k)|^2,
$$
for $\psi \in H^1$,  we have 
$$
\align
N^5  \G^N(\psi) +N \I^N(\psi) 
&  = \sum\limits_{|k| \leq m}\[ N^4|e^{2\pi i k/N} - 2 
+ e^{-2\pi i k/N}|^2  + 1\] |\hat{\psi}_N(k)|^2\\
& = \sum\limits_{|k| \leq m} 
\sigma_N^2(k) |\hat{\psi}_N(k)|^2.
\endalign
$$
Therefore,  the measures $d\, \gamma^N(\psi,\psib)$ can be written as
\footnote"*"{The sign $\asymp$ means up to unessential constant real factor.}
$$
\asymp  e^{- {1\over 2} \sum\limits_{|k|\leq m} \sigma^2_N(k) 
|\hat{\psi}(k)|^2} \underset{|k|\leq m}\to\bigwedge i\d \psih(k) 
\wedge \d \psihb(k).
$$
Similarly,  for $d\gamma(\psi,\psib)$  
$$
\asymp  
e^{-{1\over 2} \sum\limits_{k} \sigma^2(k) |\psih(k)|^2} 
\underset{k}\to\bigwedge\;  i \d \psih(k) \wedge \d  \psihb(k).
$$ 
For any fixed $k$,   
$$
\sigma_N^2(k) = N^4|e^{2\pi i k/N} -2 + e^{-2\pi i k/N}|^2 + 1 
\longrightarrow 16 \pi^4 k^4 +1 =\sigma^2(k),
$$
as $N\rightarrow \infty$, whence the convergence of measures  on 
finite-dimentional subspaces generated by the Fourier harmonics.

To prove tightness of the measures $\gamma^N$,  we introduce a "brick" 
$B_a\equiv\{\psi: |\psih(k)|\leq a_k,\;k \in \Z\}$, where $a_k\geq 0$. 
A brick is compact in $H^s$ if and only if 
$$
\sum\limits_{k} (1+k^2)^s a_k^2 <  \infty.
$$
Let us estimate
$$
\align
\gamma^N(B_a)= & \prod\limits_{|k|\leq m} {\sigma_N^2\over 2\pi} 
\int\limits_{|\psih(k)|^2 \leq a_k^2} e^{-{\sigma_N^2\over 2}|\psih(k)|^2} 
{i\over 2}\,  \d\psih(k)\wedge \d \psihb(k)\\
=  & \prod\limits_{|k|\leq m} \[ 1- e^{-{\sigma_N^2(k)a_k^2\over 2}}\].
\endalign
$$
The inequalities $(1-b_1)\times \hdots \times(1-b_n) \geq 1- b_1 -\hdots -b_n, 
\; b_k\geq 0$,  and $e^{-x} \leq p\, ! x^{-p}, \, x \geq 0$,  imply
$$
\align
\gamma^N(B_a)\geq &  1-\sum\limits_{|k|\leq m} 
e^{-{\sigma_N^2(k) a_k^2\over 2}}\\
\geq & 1- 2^p p\, ! \sum\limits_{|k|\leq m} {1\over \sigma_N^{2p} (k) a_k^{2p}}.
\endalign
$$ 
Now we obtain an estimate for $\sigma_N^2(k)=4N^4(\cos{2\pi k\over N}-1)^2 +1, 
\; |k|\leq m$. Obviously,
$$
a' x^4 \leq (\cos x -1)^2 \leq b' x^4\quad\quad \quad \text{for}\quad 
-\pi \leq x \leq \pi.
$$
Therefore,
$$
ak^4 +1 \leq \sigma_N^2(k) \leq bk^4 +1.
$$
This,  together with 
$$
a_k^2={C\over (|k|+1)^\alpha},\quad\quad\quad C>0,
$$
implies
$$
\gamma^N(B_a)\geq  1 - {2^p p\,!\over C^p} \sum\limits_{k} {(|k|+1)^{\alpha p}
\over (ak^4 +1)^p}.
$$
Pick any $s <  {3\over 2}$. Then $ 2s +1 <4$. Pick any $\alpha$ such that 
$2s + 1<\alpha< 4$. Then the bricks  $B_a$ are  compact in $H^s$.  
Now pick $p$  such that $(4-\alpha)p>1$. Then the sum in the last estimate  
converges. Chose $C$  so large as  to make $\gamma^N(B_a)$ arbitrary 
close to 1.
\qed
\enddemo

\proclaim{Lemma 7} (i). For any $\psi\in H^1$ the functionals
$$
\chi_K^N(\psi) {1\over \D_N(\psi)}e^{-{1\over 2} 
\[- N \I^N(\psi) +N^5\R^N(\psi)\] } = \chi_K^N(\psi) {1\over \D_N(\psi)} 
e^{-{1\over 2} \W^N(\psi)}
$$
converge  to
$$
\chi_K(\psi) e^{-{1\over 2}\int-|\psi|^2 +2 |\psi|^6 + 
6|\psi'|^2 |\psi|^2 +(\psib' 
\psi + \psi' \psib)^2} = \chi_K(\psi) e^{-{1\over 2} \W(\psi)} <\sqrt {e},
$$
as $N\rightarrow \infty$.

(ii). 
$$
1 \geq \D_N(\psi) \geq e^{-{K\over N}}
$$
and 
$$
\W^N(\psi) \geq - c_4(K),
$$
uniformly for all $N$, provided $N\H_1^N\leq K,\; -N^3\H_3^N \leq K$.
\endproclaim
\demo\nofrills{Proof.\usualspace} (i). The nonlinear term 
$$
\align
R^N(\psi_n,\psib_n)= 
&-\quad \sum(\psi_{n+1}\psib_{n-1} + \psib_{n+1} \psi_{n-1}) |\psi_n|^2\\
&-{1\over 2} \sum (\psi_n\psib_{n-1})^2 + (\psib_n \psi_{n-1})^2 \\
&- 6 \sum |\psi_n|^2 + 12 N I_0
\endalign
$$
can be expressed as 
$$
\align
R^N  (\psi_n, \psib_n)=&\;\;\\
+ & \sum \[(\psi_{n+1} -\psi_n)(\psib_n-\psib_{n-1}) + 
       (\psib_{n+1} -\psib_n)(\psi_n-\psi_{n-1})\] |\psi_n|^2      \tag A \\
-& \sum \[(\psi_{n+1} + \psi_{n-1}-2 \psi_n)\psib_n + 
          (\psib_{n+1} + \psib_{n-1}-2 \psib_n)\psi_n\] |\psi_n|^2  \tag B \\
-{1\over 2} &\sum \psi_n^2(\psib_{n-1} -\psib_{n})^2 + 
                 \psib_n^2(\psi_{n-1} -\psi_{n})^2 \tag C\\
-& \sum \psi^2_n \psib_n (\psib_{n-1}- \psib_n) + 
           \psib^2_n \psi_n (\psi_{n-1}- \psi_n)   \tag D \\
-3 & \sum |\psi_n|^4 - 6 \sum |\psi_n|^2 +12 N I_0.  \tag E
\endalign
$$
As in  the continuous case, if one counts the difference of $\psi$'s in two  
neighboring points $(\psi_{n+1}- \psi_{n})$ and the function $\psi_n$ itself 
of weight 1, then the terms A, B and C are isobaric polynomials of degree 6. 
The term D is isobaric polynomial of degree 5.

The term $A$ can be reduced to the form
$$
A(\psi_n,\psib_n) = \sum |\psi_{n+1} - \psi_{n-1}|^2 |\psi_n|^2 - 
\sum \{ |\psi_{n+1} - \psi_n|^2 + |\psi_{n-1} - \psi_n|^2 \} |\psi_n|^2.
$$
We transform the terms  B and D to a more convinient form also: 
$$\align
B(\psi_n,\psib_n)=&  \sum(\psi_{n+1} - \psi_n)^2 \psib_n^2 + 
                       (\psib_{n+1} - \psib_n)^2 \psi_n^2 \\
  +  & \sum |\psi_{n+1} - \psi_n|^2 \[ \psi_{n+1}(\psib_n + \psib_{n+1}) +
                                    \psib_{n+1}(\psi_n + \psi_{n+1})\] \\
=& \sum \[ (\psi_{n+1} - \psi_n)\psib_n +(\psib_{n+1}- \psib_n) \psi_n \]^2 \\
+ & 2 \sum |\psi_{n+1} - \psi_{n}|^2 |\psi_{n+1}|^2\\
+ & \sum |\psi_{n+1}- \psi_n|^2 \[(\psi_{n+1}-\psi_n) \psib_{n} + (\psib_{n+1}- 
\psib_n)\psi_n\]
\endalign
$$
and
$$\align
D(\psi_n,\psib_n)={1\over 2} & \sum \psi_n^2(\psib_n -\psib_{n-1})^2 + 
\psib_n^2(\psi_n-\psi_{n-1})^2\\
+ 2 & \sum |\psi_{n+1} - \psi_n|^2 |\psi_n|^2\\
+ {1\over 2}&  \sum |\psi_{n+1} - \psi_n|^2 \[ \psib_n(\psi_{n+1} - \psi_{n}) + 
\psi_n(\psib_{n+1} - \psib_{n})\] \\
-{1\over 2} &\sum |\psi_n|^2 \[|\psi_{n+1}-\psi_n|^2-|\psi_{n-1}-\psi_n|^2\].
\endalign
$$
The first term in the formula is C with the opposite sign. Finally,
$$
\align
R(\psi_n,\psib_n) =\quad \quad  & \\
- 3  & \sum |\psi_n|^4 - 6 \sum |\psi_n|^2 +12 N I_0 \tag 1 \\ 
+ &   \sum  |\psi_{n+1} - \psi_{n-1}|^2 |\psi_n|^2 \tag 2 \\
+ {3\over 2} & \sum |\psi_{n-1} - \psi_n|^2 |\psi_n|^2 \tag 3 \\
+{1\over 2} & \sum |\psi_{n+1} - \psi_n|^2 |\psi_n|^2 \tag 4 \\
+& \sum \[(\psi_{n+1} - \psi_n) \psib_n + (\psib_{n+1} - \psib_n) \psi_n\]^2
\tag 5\\
+  {3\over 2}& \sum|\psi_{n+1}- \psi_n|^2 \[ \psib_n(\psi_{n+1} - \psi_n) +
\psi_n(\psib_{n+1} - \psib_n)\] \tag 6         
\endalign
$$
The advantage of such a representation is that the terms (1)-(5) are 
nonnegative. The term  (6) is the only term which is not sign-definite. 
This term vanishes when $N\rightarrow \infty$.
Finally,  as $N\rightarrow \infty$
$$
\align
& N^5 \R^N(\psi)\longrightarrow \int\limits_{0}^{1} 2|\psi|^6 + 6|\psi'|^2 
|\psi|^2 + \(\psib'\psi + \psi' \psib\)^2,\\
& N \I^N(\psi) \longrightarrow  \int\limits_{0}^{1} - |\psi|^2\\
&\D_N(\psi)\longrightarrow 1.
\endalign
$$

(ii).  From the definition $H_1=N(I_0+\bar{I}_0)=-\log D$ we have 
$
\H_1^N (\psi)= -\log \D_N(\psi).
$
The inequality $N\H_1^N \leq K$ implies, that $1 \geq \D_N \geq 
e^{-{K\over N}}$.

To obtain the lower bound for $R^N(\psi_n,\psib_n)$, we prove first  that 
$$
-3\sum |\psi_n|^4 - 6 \sum |\psi_n|^2 + 12 N I_0 \geq 2\sum |\psi_n|^6. \tag 7
$$
Indeed, for $0 < x < 1$
$$
-\log(1-x) \geq x + {x^2\over 2} + {x^3\over 3}.
$$
Then, 
$$
-6 \log (1-|\psi_n|^2) \geq 6 |\psi_n|^2 + 3 |\psi_n|^4 + 2|\psi_n|^6.
$$
This and $12NI_0= -6 \log \prod (1-|\psi_n|^2)$ imply the inequality (7). 
Therefore, the term (1) is nonegative. 

We have just one term (6) which is not sign-definite
$$
{3\over 2}\sum |\psi_{n+1}- \psi_n|^2 \[ \psi_n \psib_{n+1} + 
\psib_{n} \psi_{n+1} - 2 |\psi_n|^2\].
$$
Using the identity
$$
|\psi_{n+1}- \psi_n|^2 -|\psi_{n+1}|^2 - |\psi_n|^2 = \psi_{n+1} \psib_n + 
\psib_{n+1} \psi_n 
$$
and adding to it the two positive terms (3) and (4), we have 
$$
\align
{3\over 2}  & \sum |\psi_{n+1} -\psi_{n}|^2 \( \psi_n \psib_{n+1} + \psib_{n} 
\psi_{n+1} - 2|\psi_n|^2\)\\
+ {3\over 2}& \sum |\psi_{n-1}- \psi_n|^2 |\psi_n|^2\\ 
+  {1\over 2} & \sum |\psi_{n+1}-\psi_n|^2 |\psi_n|^2\\
 \geq&  - 4\sum |\psi_{n+1} - \psi_n|^2 |\psi_n|^2.
\endalign
$$
Therefore,
$$
R(\psi_n,\psib_n)\geq - 4 \sum |\psi_{n+1} - \psi_n|^2 |\psi_n|^2.
$$
and
$$
N^5 \R(\psi) \geq -4 \sum N |\psi({n+1/ N}) - \psi({n/ N})|^2 
|\psi({n/ N})|^2.
$$
Now we use the constraint $-N\H_3^N \leq K$ and the explicit formula for $H_3$:
$$
-H_3^N(\psi_n,\psib_n)= \sum |\psi_{n+1} - \psi_n|^2 -2|\psi_n|^2 - 2\log(1-
|\psi_n|^2).
$$
The estimate $-\log(1-x) \geq x +{x^2\over 2}$  for $0\leq x <1$ implies  
$$
-2 \log (1-|\psi_n|^2) \geq 2 |\psi_n|^2 + |\psi_n|^4.
$$
Therefore,
$$
-H_3^N(\psi_n,\psib_n)\geq \sum |\psi_{n+1} - \psi_n|^2 + |\psi_n|^4.
$$
Finally,
$$
K\geq -N^3 \H_3^N(\psi) \geq \sum N|\psi(n+1/N) - \psi(n/N)|^2 + N^{-1}
|\psi(n/N)|^4.
$$
This inequality  also  provides the estimate for \linebreak 
$\max |\psi(n/N)|$\footnote"*"{This is similar to the continuum case when 
the $H^1$-norm provides a bound in the sup-norm.}. 
Indeed, 
$$
{1\over N} \sum |\psi(n/N)|^4 \leq K ,
$$
implies that there exists $n''$ such that $|\psi(n''/N)|\leq K^{1/4}$. 
Then,  for any $n'$, 
$$
\psi(n'/N) = \sum\limits_{n'\leq n < n''} \[ \psi(n+1/N) - \psi(n/N)\] 
+ \psi(n''/N)
$$
and,  by Schwartz's inquality,
$$
\align
|\psi(n'/N)|& \leq \sum |\psi(n+1/N) - \psi(n/N)| {\sqrt{N}\over \sqrt{N}} 
+|\psi(n''/N)|\\
&\leq  \sqrt{\sum N |\psi(n+1/N) - \psi(n/N)|^2} + | \psi(n''/N)|\\
& \leq c_2(K).
\endalign
$$
Therefore,
$$
\align
N^5 \R^N(\psi) & \geq - 4 \max |\psi(n/N)|^2 \sum N|\psi(n+1/N) - \psi(n/N)|^2\\
             & \geq - 4 c_2^2(K) K = -c_3(K).
\endalign
$$
Also,
$$
N\I^N(\psi)= \sum N^{-1} |\psi(n/N)|^2 \leq c^2_2(K).
$$
Finally,
$$
\W^N(\psi)=-N\I^N(\psi) +N^5 \R^N(\psi) \geq - c_2^2(K) - c_3(K)=-c_4(K).
$$
\qed
\enddemo

\proclaim{Lemma 8} (i) The  probability measures 
$$d\mu_K^N = {1\over \Xi_K^N} \chi_K^N(\psi) 
{1\over \D_N(\psi)}e^{-{1\over 2} \W^N(\psi)}
d\gamma^N(\psi,\psib),
$$
with
$$
\Xi_K^N\equiv \int\limits_{\M} \chi_K^N(\psi) {1\over \D_N(\psi)}
e^{-{1\over 2} \W^N(\psi)} d\gamma^N(\psi,\psib)
$$
converge weakly in $H^s, 1\leq s < {3\over 2}$ as $N\rightarrow \infty$ to 
the probability measure 
$$d\mu_K={1\over \Xi_K} 
\chi_K(\psi) e^{-{1\over 2}\W(\psi)} d\gamma(\psi,\psib),
$$
with 
$$
\Xi_K\equiv
\int\limits_{\M}\chi_K(\psi) e^{-{1\over 2} \W(\psi)} d\gamma(\psi,\psib).
$$

(ii). The measure $d\mu_K$ converges in $H^s,\; 1\leq s <{3\over 2}$ as 
$K\rightarrow \infty$ to the probability measure 
$$
d\mu={1\over \Xi} e^{-{1\over 2} \W(\psi)} d\gamma(\psi,\psib),
$$ 
with
$$
\Xi\equiv \int\limits_{\M} e^{-{1\over 2} \W(\psi)} d\gamma (\psi,\psib).
$$

\endproclaim
\demo\nofrills{Proof.\usualspace} (i). Take $\M=H^s\times H^s,\; 
1\leq s \leq {3\over 2}$ and $f$ bounded  and continuous. We will prove 
that 
$$
\int\limits_{\M} f \chi_K^N {1\over \D_N(\psi)}\,  e^{-{1\over 2} \W^N(\psi)} 
d\gamma^N(\psi,\psib) \longrightarrow \int\limits_{\M} f \chi_K e^{-{1\over 2}
\W} d\gamma(\psi,\psib), \tag 8
$$
as $N\rightarrow \infty$. 
This implies that $\Xi_K^N\rightarrow \Xi_K$ as $N\rightarrow \infty$. 
Then,  from (8), we have  
$$
\align
\int f \, d\mu_K^N= & {1\over \Xi_K^N}\int f\, \chi_K^N {1\over \D_N(\psi)}
e^{-{1\over 2}\W^N} d\gamma^N(\psi,\psib)\\ 
& \longrightarrow {1\over \Xi_K} \int f\chi_K e^{-{1\over 2}\W} d \gamma(\psi,\psib)\times 
{\Xi_K\over \Xi_K^N} = 
\int f d\mu_K \times {\Xi_K\over \Xi_K^N},
\endalign
$$ 
as $N\rightarrow \infty$  and weak convergence is proved. 

To begin with,
$$
\align
 \int f \chi_K^N & {1\over \D_N(\psi)} e^{-{1\over 2} \W^N(\psi)} 
d\gamma^N(\psi,\psib)\\
&=\int f\( {1\over \D_N(\psi)} -1\) \chi_K^N e^{-{1\over 2}\W^N} d\gamma^N + 
\int f \chi_K^N e^{-{1\over 2}\W^N} d\gamma^N(\psi,\psib).
\endalign
$$
The first term can be estimated using the inequality of Lemma 8, item  (ii):
$$
\therefore\; \leq  \|f\|_{\infty}\,  e^{{1\over 2}c_4(K)} \int\limits_{\M} 
\left|{1\over \D_N(\psi)} -1\right|\chi_K^N d\gamma^N(\psi, \psib) = o(1),
$$
as $N\rightarrow \infty$. As to  the second term,  
$$
\align
  \int f \chi_K^N e^{- {1\over 2}\W^N} d\gamma^N(\psi,\psib)
 =& \int f \chi_K e^{-{1\over 2} \W} d\gamma^N(\psi,\psib)  \\
 +& \int f \chi_K^N \(e^{-{1\over 2}\W^N} - e^{-{1\over 2} \W}\) 
     d\gamma^N(\psi,\psib) \tag 9  \\
+& \int f \(\chi_K^N -\chi_K\) e^{-{1\over 2}\W} d\gamma^N(\psi,\psib). \tag 10
\endalign
$$ 
The first integral on the right  converges to 
$$
\int f \chi_K e^{-{1\over 2} \W} d \gamma(\psi,\psib). 
$$
The other two integrals vanish as $N\rightarrow \infty$. 
Indeed,  (9) can be overestimated by  
$$
{\|f\|_{\infty}\over 2} e^{{c_4(K)\over 2}} \int \chi_K^N |\W^N - 
\W| d\, \gamma^N(\psi, \psib). \tag 11
$$
Using the explicit expressions for $\W^N$ and $\W$, we have 
$$
\align
\W^N -  \W & = \\
- &  \sum N^{-1} |\psi({n/N})|^2 + \int |\psi|^2 dx \tag 12\\
 - 3& \sum N |\psi({n/N})|^4 - 6\sum N^{3} 
|\psi({n/ N})|^2 + 12N^6 \I_0(\psi) - 2\int  |\psi|^6dx \tag 13\\
+ &  \sum N|\psi(n+1/N) - \psi(n-1/N)|^2 |\psi(n/N)|^2 - 4 \int 
|\psi'|^2|\psi|^2 dx \tag 14\\
 + {3\over 2}&  \sum N |\psi(n-1/N) -\psi(n/N)|^2 |\psi(n/N)|^2 - {3\over 2} 
\int |\psi'|^2 |\psi|^2 dx \tag 15\\
+{1\over 2}&\sum N|\psi(n+1/N) - \psi(n/N)|^2 |\psi(n/N)|^2 - {1\over 2} \int 
|\psi'|^2 |\psi|^2 dx \tag 16 \\ 
+ & \sum N [ (\psi(n+1/N) - \psi(n/N)) \psib(n/N) + \tag 17 \\
&+ (\psib(n+1/N)-\psib(n/N))\psi(n/N)]^2-\int[\psi'\psib+\psib\psi']^2 dx \\  
+ {3\over 2} &  \sum N |\psi(n+1/N) - \psi(n/N)|^2 [\psib(n/N)(\psi(n+1/N) - 
\psi(n/N)) + \tag 18 \\
& + \psi(n/N)(\psib(n+1/N) - \psib(n/N))]. 
\endalign 
$$
Here, each term contributes nothing  when $N\rightarrow \infty$; for example, 
substituting (12) into (11):
$$
\align
\leq & \int_{\M} \left|\sum N^{-1} |\psi(n/N)|^2 - \sum\int_{n/N}^{n+1/N} 
|\psi|^2 d x\right|d\gamma^N(\psi,\psib)\\
\leq & \sum\int_{\M}\left| N^{-1}|\psi(n/N)|^2 - \int_{n/N}^{n+1/N} |\psi|^2 dx 
\right| d\gamma^N(\psi,\psib)\\
\leq & N\int_{\M} \left |\int_{0}^{1/N} |\psi(x)|^2 -|\psi(0)|^2 
dx \right |d\gamma^N(\psi,\psib)\\
\leq & N\int\limits_{0}^{1/N} \int_{\M} \left| |\psi(x)|^2 - |\psi(0)|^2
\right| d\gamma^N 
(\psi,\psib) dx.
\endalign
$$
Now, by Schwartz, for $ 0 \leq |x| \leq {1/N}$, we have  
$$
\align
\(\int_{\M} \left| |\psi(x)|^2-|\psi(0)|^2\right|d\gamma^N\)^2 
\leq & \int_{\M} \(|\psi(x)|^2 - |\psi(0)|^2\)^2 d\gamma^N\\
= & \int_{\M} |\psi(x)|^4-2|\psi(x)|^2|\psi(0)|^2 + |\psi(0)|^4 
d\gamma^N \\
=& 2\int_{\M} |\psi(0)|^4 - |\psi(x)|^2 |\psi(0)|^2 d\gamma^N
\endalign
$$
The stationary Gaussian measure $d\gamma^N$ has the spectral representation  
$$
\psi_N(x)= \sum \Sb |k| \leq m \endSb e^{2\pi i k x} \hat{\psi}(k)
$$
where $\hat{\psi}$ are complex isotropic  Gaussian variables 
$E^N|\hat{\psi}(k)|^2 = \sigma_N^{-2}(k)$.   Now 
$$
|\psi(0)|^2 |\psi(x)|^2 = \sum \Sb k_1,k_2\\ k_3, k_4 \endSb 
e^{2\pi i x (k_1- k_2)} \hat{\psi}(k_1) \overline {\hat{\psi}}(k_2) 
\hat{\psi}(k_3) \overline{\hat{\psi}}(k_4).
$$
The terms in the last sum  contributing into the expectation $E^N$ are 
$$
\xalignat 3
k_1& = k_2    &   k_1& =k_4    &  k_1& =k_2=k_3=k_4.\\
k_3& = k_4    &   k_3& =k_2    & \\
k_1&\neq k_3  &   k_1&\neq k_3
\endxalignat
$$ 
Now 
$$
\multline
E^N  |\psi(0)|^2 |\psi(x)|^2 = \sum \Sb k_1\neq k_3 \endSb 
E^N|\psih(k_1)|^2 E^N|\psih(k_3)|^2\\  
+  \sum \Sb  k_1\neq k_2 \endSb e^{2\pi i x(k_1-k_2)} 
E^N|\psih(k_1)|^2 
E^N|\psih(k_2)|^2 + \sum \Sb k_1 \endSb E^N|\psih(k_1)|^4,
\endmultline
$$
So,
$$
\align
E^N|\psi(0)|^4 -  E^N & |\psi(x)|^2 |\psi(0)|^2\\
& =\sum \Sb k_1 \neq k_2 \endSb \[ 1- e^{2\pi i x(k_1-k_2)}\] 
E^N|\psih(k_1)|^2 E^N|\psih(k_2)|^2 \\
& =\sum \Sb |k_i| \leq m \endSb  |1- e^{2\pi i x(k_1-k_2)}| 
\sigma_N^{-2}(k_1) \sigma_N^{-2}(k_2).
\endalign
$$
Using the estimate $a k^4 + 1 \leq \sigma_N^2(k)$ of Lemma 7, uniformly in 
$0\leq x \leq 1/N$, the last sum can be overestimated as
$$
\therefore\; \leq \sum \Sb k_i \in \, \Z^1 \endSb 
\left |1- e^{2\pi i x(k_1- k_2)}\right| 
{1\over ak_1^4 + 1} {1\over ak_2^4 +1}= o(1),
$$
as $N\rightarrow \infty$. This implies that the term (12) contributes nothing  
when $N\rightarrow \infty$.

To estimate the contribution of (13),  expand the $\log$ in $\I_0$: 
$$
\align
12N^6\I_0(\psi)=&-6N^5\log\prod(1-N^{-2}|\psi(n/N)|^2)\\
  =& 6N^5[ N^{-2} |\psi(n/N)|^2 + {N^{-4}\over 2}|\psi(n/N)|^4 
+{N^{-6}\over 3} |\psi(n/N)|^6\\ 
+& {f^{(iv)}\over 4!}N^{-8} |\psi(n/N)|^8 ]
\endalign
$$
where $f(x) = -\log(1-x)$ and the fourth derivative $f^{(iv)}= 3!(1-x)^{-4}$ 
in the remainder term is bounded since $|\psi(n/N)|\leq c_2(K)$ and 
$N^{-2}|\psi(n/N)|^2\leq \theta < 1$. 

Now for  (13), we have 
$$
2\sum N^{-1} |\psi(n/N)|^6 - 2\int\limits_{0}^{1} |\psi|^6 dx
+  {6\over 4!} N^{-3} \sum f^{(iv)} |\psi(n/N)|^6.
$$
The part with the the fourth derivative can be overestimated by 
$$
\therefore\; \leq {6\over 4!}N^{-3} {3!\over (1-\theta)^4} c_2^6(K) N 
=o(1),
$$
as $N\rightarrow \infty$.  
The contribution to (11) of  first two terms can be overestimated by
$$
\align
\therefore &\leq 2 \left|\int_{\M} \sum N^{-1} |\psi(n/N)|^6 - 
\int_{n/N}^{n+1/N} |\psi|^6 dx \right | d\gamma^N\\
& \leq 2 N \int_{0}^{1/N} \int_{\M}\left||\psi(x)|^6-|\psi(0)|^6\right|  
d\gamma^N  dx. 
\endalign
$$
Again, by Schwarz,  
$$
\align
& \(\int_{\M}\left| |\psi(x)|^6 - |\psi(0)|^6\right| d \gamma^N\)^2\\
\leq 2 & \int_{\M} |\psi(0)|^{12} - |\psi(0)|^6 |\psi(x)|^6 d\gamma^N = o(1),
\endalign
$$
as $N\rightarrow \infty$,  uniformly in $0\le x < 1/N$. This implies that  the 
term (13) makes no contribution when $N\rightarrow \infty$. The estimation 
of  (14-18) is similar. 

(10) can be overestimated  by 
$$
\|f\|_{\infty} \int |\chi_K^N-\chi_K|d\gamma^N. \tag 19
$$
Now, we have
$$
\chi_K^N-\chi_K= \[h(N\H_1^N)- h(\H_1)\] h(-N^3\H_3^N) 
                +\[h(-N\H_3^N) -h(\H_3)\] h(\H_1). 
$$
Therefore, (19) can be overestimated by 
$$
\align
\therefore\;\leq& \|f\|_{\infty}\int|h(N\H_1^N)-h(\H_1)|h(-N^3\H_3^N)d\gamma^N
\tag 20\\
+ & \|f\|_{\infty} \int | h(-N^3\H_3^N) - h(\H_3)|h(\H_1) d\gamma^N. \tag 21
\endalign
$$
To estimate (20) we note that $|\psi(n/N)|\leq c_2(K)$  on the \break 
support of $h(-N^3\H^N_3)$ and that  
$$
\align
N\H_1^N& = -N\log \prod (1-N^{-2}|\psi(n/N)|^2)\\
        &=N^{-1}\sum |\psi(n/N)|^2 
+{N^{-3}\over 2!} \sum f^{(ii)} |\psi(n/N)|^4.
\endalign
$$
Therefore,
$$
\align
(20)  \leq \|f\|_{\infty} \|h'\|_{\infty} & \int_{|\psi(n/N)|\leq 
c_2(K)} | N\H^N_1-\H_1|\; d\gamma^N\\
\leq  \|f\|_{\infty} \|h'\|_{\infty}& \int_{|\psi(n/N)|\leq c_2(K)} \left| 
N^{-1} \sum |\psi(n/N)|^2 -\int_{0}^{1} |\psi|^2\, dx \right| d\gamma^N \\
+& \|f\|_{\infty} \|h'\|_{\infty} \int_{|\psi(n/N)|\leq c_2(K)} 
\left| \sum {f^{(ii)}\over 2!} N^{-3} |\psi(n/N)|^4\right|d\gamma^N.
\endalign
$$
The first term vanishes as in the estimate for the term (20) and the 
second term vanishes as in the estimate for the term (21). 

To estimate the term (21), note that for any $\epsilon > 0$, 
$$
\gamma^N\( \|\psi\|_{\infty} \geq C \) \geq 1- \epsilon,
$$
for all $N$ provided  $C$ is sufficiently large. Then,  
$$
(21) \leq 2\epsilon + \int_{\|\psi\|_{\infty} \leq C} |h(-N^3\H_3^N) -
h(\H_3)|h(\H_1) d \gamma^N, 
$$
in which the integral vanishes   for $N\rightarrow \infty$ as for (20).

(iv). For any bounded continuous $f$
$$
\int f \chi_K(\psi) e^{-{1\over 2} \W(\psi)} d\gamma(\psi,\psib) \rightarrow
\int f  e^{-{1\over 2}\W(\psi)} d \gamma(\psi,\psib),
$$
as $K\rightarrow \infty$, by the bounded convergence theorem. This 
implies $\Xi_K \rightarrow \Xi$ as $K \rightarrow \infty$. Now  the last 
statement is proved.
\qed
\enddemo

\subhead 10. Tightness of measures $d\m_K^N$\endsubhead We assume that the 
initial data $\psi_N(x)$ has distribution $d\mu_N^K$. 
The solutions of initial  value problem  $\psi(x,t)=e^{t\X_3^N}\psi_N(x),$ 
$ (x,t)\in \T\times [0,T]$,  are realisations of the 
complex random field with distribution $d\m_K^N$. 
To prove tightness of measures  we need 
\proclaim{Lemma 9} (Kolmogorov---\v Centzov). \cite{Ku}. 
Let the family of complex 
continuous random fields $\phi(x,t),\; (x,t) \in \T\times[0,T]$,  with 
distributions $dm^N$ satisfy 
$$
E^N|\phi(x,t)|^\gamma \leq C \tag 1
$$
and
$$
E^N|\phi(x_1,t_1) - \phi(x_2,t_2)|^\gamma \leq C\(|x_1-x_2|^{\alpha_1}+ 
|t_1-t_2|^{\alpha_2}\), \tag 2
$$
where $\alpha^{-1}_1 +\alpha^{-1}_2 < 1$ and $\gamma > 0$. 
Then the family of measures $dm^N$ is tight with respect to the weak 
topology of $C\(\phi:\T\times[0,T]\rightarrow \C^1\)$ 
\endproclaim 
Since realisations of the field are continuously differentiable in the spatial 
variable $x$,  we consider the random fields $\psi'(x,t)$.  
The main result of this section is 
\proclaim{Lemma 10} For any $n=1,2,\hdots,$: 
$$
\align 
E_K^N&|\psi'(x,t)|^{2n} \leq c(n,K) \tag 3\\ 
E_K^N&|\psi'(x_1,t_1)-\psi'(x_2,t_2)|^{2n}\leq c(n,K)\(|x_1-x_2|^{n} + 
|t_1-t_2|^{n/2}\) \tag 4
\endalign
$$
where $E_K^N$ is expectation with respected to the measure $d\m_K^N$.
\endproclaim
This, together with Lemma 9, implies tightness of the measures  $d\m'{}^N_K$. 
Tightness of the family $d\m_K^N$ can be proved along the same lines. The 
distribution of the limiting stationary random field we denote by $d\m_K$. 
It follows from the  previous section,  that the measure $d\m_K$ has 
marginal distributions $d\mu_K$. In order to establish the H\"{o}lder 
continuity 
of the spatial derivatives of the random field we need another 
\proclaim{Lemma 11} (Kolmogorov--\v Centzov). \cite{Ku}. 
Let $\phi(x,t),\; (x,t)\in 
\T\times[0,T] $,  be a complex random field. Assume that there exist positive 
constants $\gamma,\; C,\; \alpha_1$ and  $\alpha_2$ with $\alpha^{-1}_1+ 
\alpha_2^{-1}< 1$ satisfying 
$$
E|\phi(x_1,t_1)-\phi(x_2,t_2)|^\gamma \leq C\(|x_1-x_2|^{\alpha_1} 
+ |t_1-t_2|^{\alpha_2} \). 
$$
Then the random field $\psi(x,t)$ has a continuous modification. 
Moreover $c_0\equiv (1- \alpha^{-1}_1 -\alpha^{-1}_2)/\gamma$, 
if $\beta_1$ and $\beta_2$ are positive numbers less then $\alpha_1c_0$ or 
$\alpha_2c_0$ respectively, then there exists a 
positive random constant $C$ with $E C^\gamma <\infty$ such that 
$$
|\phi(x_1,t_1)-\phi(x_2,t_2)|< C\(|x_1-x_2|^{\beta_1} + |t_1-t_2|^{\beta_2}\).
$$
\endproclaim
By Fatou's lemma, passing to the limit $N\rightarrow \infty$ in the inequality 
(4), we have 
$$
E_K|\psi'(x_1,t_1)-\psi'(x_2,t_2)|^{2n} \leq C(n) \(|x_1-x_2|^{n}+
|t_1-t_2|^{n/2}\)  
$$
Now Lemma 11  implies that the measure $d\m'_K$ is supported on  paths 
satisfying 
$$
 |\psi'(x_1,t_1) - \psi'(x_2,t_2)|\leq\(|x_1-x_2|^{1/2-} + |t_1-t_2|^{1/4-}\).
$$
One can prove, using methods of \cite{V}, that the H\"{o}lder exponents 
$1/2-$ and $1/4-$ are optimal.

\demo\nofrills{Proof of Lemma 10.\usualspace}
To obtain (3), observe that by the invariance of $d\mu_K^N$
$$
\align
E_K^N |\psi'(x,t)|^{2n}& =E_K^N|\psi'(x,0)|^{2n}  \\
&\leq {e^{K\over N} e^{c_4\over 2}\over \underset N\to \inf \; \Xi_K^N}
\int\limits_{\M}|\psi'(x,0)|^{2n} d\gamma^N \quad \quad \quad
\text{by Lemma}\; 7
\endalign
$$
Due to Lemma 8,  $\Xi_K^N\rightarrow \Xi_K> 0 $,  as $N\rightarrow \infty$. 
Using  the Gaussian character of the measure $d\gamma^N$ the last integral 
can be overestimated  by  
$$
\leq c_1(K,n) \[\; \int\limits_{\M} |\psi'(0)|^{2}d\gamma^N \]^n=
c_2(K,n) \[\sum\limits_{|k|\leq m} k^2\sigma_N^{-2}(k)\]^n
$$
Now the estimate $ak^4 + 1 \leq \sigma_N^2(k)$ (see proof of Lemma 6) implies 
(3).

To prove  (4),   write the differential equation for the 
flow on $\M_N$ as in
$$
\align
i{\partial \psi(x)\over \partial t} =-N^2(\psi(x+{1\over N})+&\psi(x-{1\over N})
- 2\psi(x)) \\
+& |\psi(x)|^2 (\psi(x-{1\over N}) +\psi(x+{1\over N})),
\endalign
$$
where $\psi(x)=\psi_N(x,t) \in \M_N,\;\; x\in \T_N$;  equivalently,     
$$
\align
e^{t\X_3^N} \[\psi(\bullet)\]& (x)=  e^{it\Delta_N}\[\psi(\bullet)\](x)\\
- & i\int_{0}^{t} e^{i(t-s)\Delta_N} \[|\psi(\bullet,s)|^2
\(\psi(\bullet+{1\over N},s) +\psi(\bullet -  {1\over N},s)\)\]_N(x) ds, \tag 5
\endalign
$$
where 
$$
e^{it\Delta_N}[\psi(\bullet)](x)\equiv\sum\limits_{|k|\leq m} e^{2\pi i k x}
e^{it \Delta(k,N)} \psih(k),\quad \quad \quad \Delta(k,N)=N^2(\omega^k +
\omega^{-k}-2)
$$
and
$$
\psi(x)=\sum\limits_{|k|\leq m} e^{2\pi i  kx} \psih(k). 
$$
Since (5) holds for all $x\in \T_N$ with left and right   
being polynomials, then (5) holds for all $x\in \T$. 

First,  we derive the estimate for spatial increments. 
Much as before, 
$$
\align
E_K^N|& \psi'(x+h,t) -\psi'(x,t)|^{2n} \\
& = E_K^N|\psi'(x+h) - \psi'(x)|^{2n}\\
&\leq c_2(K,n) \[E_{\gamma^N} |\psi'(h)-\psi'(0)|^2 \]^n.
\endalign
$$
Using the spectral representation of the Gaussian measure $\gamma^N$ in the 
form  
$$
\psi(x)=\sum\limits_{|k|\leq m} e^{2\pi i k x} \psih(k), \quad \quad 
\quad \quad E_{\gamma^N} |\psih(k)|^2=\sigma_N^{-2}(k),
$$
we obtain
$$
E_{\gamma^N}|\psi'(h)-\psi'(0)|^2 =\sum\limits_{k\neq 0} 4\pi^2 k^2 
|e^{2\pi i kh}-1|^2 \sigma^{-2}_N(k).
$$
The estimate $|e^{i2\pi x}-1|\leq cx, \;\; 0\leq |x|\leq {1\over 2}$, implies  
$$
\align
& \leq c_3\sum\limits_{0 < |k| < h^{-1}/2} 
{k^4 h^2\over a k^4 +1} + c_3 \sum\limits_{h^{-1}/2 \leq \, |k|} 
{k^2\over ak^4 +1}\\
& \leq c_4h^2\sum\limits_{0<|k|<\, h^{-1}/2} 1 + c_4 
\sum\limits_{h^{-1}/2 \leq \, |k|}{1\over k^2}  \leq c_5 h.
\endalign
$$
Finally,
$$
E_K^N|\psi'(x+h,t)-\psi'(x,t)|^{2n} \leq c_6(K,n)h^n. \tag 6
$$

To obtain the estimate for temporal  increments  we use (5) and  
write\footnote"*"{$e^{it\Delta_N} \psi(x)=e^{it\Delta_N}\[\psi(\bullet)\] (x)$} 
$$
\align
E_K^N |\psi'(x,& t+h) - \psi'(x,t)|^{2n}  = E_K^N|\psi'(x,h)-\psi'(x,0)|^{2n}\\
\leq & c_7\, E_K^N\left|e^{ih\Delta_N}\psi'(x)-\psi'(x)\right|^{2n} \tag 7 \\
+ &  
c_7\; E_K^N\left|\int\limits_{0}^{h}e^{i(h-s)\Delta_N}\[|\psi(\bullet)|^2
\(\psi(\bullet + {1\over N},s)+ \psi(\bullet-{1\over N},s)\)\]'_N(x)ds 
\right|^{2n} \tag 8 
\endalign
$$

Here 
$$
(7)\quad \leq c_8\; E_{\gamma^N}|e^{ih\Delta_N}\psi'(x) -\psi'(x)|^{2n}=
c_9 \[E_{\gamma^N} |e^{ih\Delta_N}\psi'(x)-\psi'(x)|^2\]^{n}.
$$
Now, from  the spectral respresentation,  
$$
\align
E_{\gamma^N}|e^{ih\Delta_N} \psi'& (0) -\psi'(0)|^2 
= \sum\limits_{|k|\leq m} |e^{ihN^2(\omega^k+\omega^{-k}-2)} 
-1|^2 4\pi^2 k^2 \sigma_N^{-2}(k)\\
=&\sum\limits_{|k|\leq m} 
\(1- \cos hN^2(\omega^k+\omega^{-k}-2)\)^2 4\pi^2 k^2 \sigma_N^{-2}(k) \tag 9\\
+& \sum \limits_{|k|\leq m}  
\sin^2 hN^2(\omega^k+\omega^{-k}-2) 4\pi^2 k^2 \sigma_N^{-2}(k).  \tag 10 
\endalign
$$
First, we note that for $|x|\leq \pi$ 
$$
c_{10}x^2 \leq 1-\cos x \leq c_{11}x^2 \tag 11
$$
Therefore,  for $|k|\leq  m$, we have 
$$
c_{12}k^2 \leq -N^2(\omega^k+\omega^{-k}- 2) \leq c_{13} k^2. \tag 12
$$
To estimate (8), we assume that $|k|\leq h^{-1/2}\alpha$,  where 
$\alpha$ is a suitably chosen constant. Then, by (12),  
$$
\left|hN^2(\omega^k+\omega^{-k} - 2)\right|\leq \pi.
$$
Using (11) and (12), we  find
$$
\align
(9)  &\leq c_{14}  \sum \Sb |k|\leq m\\ |k|\leq h^{-1/2}\alpha \endSb
{\(hN^2(\omega^k+\omega^{-k}-2)\)^4\over k^2+1} + 
c_{15}\sum \Sb |k| \leq m\\ |k|>  h^{-1/2}\alpha \endSb {1\over k^2+1}\\
& \leq c_{16}h^4  \sum\limits_{|k|\leq h^{-1/2}\alpha} k^6+
c_{17}\sum\limits_{|k| >  h^{-1/2} 
\alpha} {1\over k^2} \leq c_{18} h^{1/2}.
\endalign
$$

The estimate of (10) is even simpler:
$$
\align
(10)\quad & \leq c_{19} \sum \Sb |k|\leq m\\ |k|\leq h^{-1/2}\endSb 
{h^2 k^4\over k^2+1} + c_{20}
\sum\Sb |k|\leq m\\|k|>  h^{-1/2} \endSb {1\over k^2+1}
\\
& \leq c_{21} h^2\sum\limits_{|k|\leq h^{-1/2}} k^2 + 
c_{22}\sum\limits_{|k|>  h^{-1/2}}
{1\over k^2}\leq c_{23} h^{1/2}.
\endalign
$$ 

Finally, for  (7) we have the estimate  
$$
E_K^N|e^{it\Delta_N}\psi'(x) - \psi'(x)|^{2n}\leq c_{24}(K,n)h^{n/2}. \tag 13
$$

To estimate (8),   observe that  
$$
\align
E_K^N& \left|\int_{0}^{h} e^{i(h-s)\Delta_N} \[|\psi(\bullet,s)|^2(\psi(\bullet+
{1\over N},s) +\psi(\bullet-{1\over N},s))\]_N'(x) ds \right|^{2n}\\
& \leq h^{2n-1}\int_{0}^{h}E_K^N\left|e^{i(h-s)\Delta_N} \[|\psi(\bullet,s)|^2(
\psi(\bullet+{1\over N},s)+\psi(\bullet -{1\over  N},s))\]_N'\right|^{2n} ds\\
&\leq h^{2n-1} \int_{0}^{h}E_K^N\left|e^{is\Delta_N}\[|\psi(\bullet,s)|^2
(\psi(\bullet +{1\over N},s)+\psi(\bullet -{1\over N},s))\]_N'\right|^{2n}ds.
\endalign
$$
We will show that,  for any $s\geq 0$,
$$
E_K^N\left|e^{it\Delta_N}\[|\psi(\bullet)|^2(\psi(\bullet +{1\over N},s) + 
\psi(\bullet -{1\over N},s))\]_N'\right|^{2n} \leq c_{25}(K,n). \tag 14
$$
Then, 
$$
(8)=E_K^N\left|\quad\hdots ditto \hdots \quad\right|^{2n} \;\leq c_{25}h^{2n}. 
\tag 15
$$ 
The estimates (13) and (15) produce 
$$
E_K^N|\psi'(x,t+h) -\psi'(x,t)|^{2n}\leq c_{26}(K,n)h^{n/2}.
$$
This  and (6) produce   (4).

To prove (14) first we note, is $\psi(x)=\sum\psih(k) e^{2\pi i  kx}$, then 
$$\psi_N(x)=\sum\limits_{|k|\leq m} \psih_N(k)e^{2\pi i kx},\quad \quad 
\text{with} 
\quad \quad  
\psih_N(k)=\sum\limits_{n\equiv k,\;  (\mod N)} \psih(n).$$ 
To simplify notation we write $\psi=\psi_N$, then 
$$
\align
|\psi(x)|^2\(\psi(x+{1\over N}) + \psi(x-{1\over N}) \)& \\
= \sum\limits_{|k_i|\leq m}& 
e^{2  \pi i (k_1+k_3-k_2)x} 2\cos {2\pi k_3\over N} 
\psih(k_1)\psihb(k_2) \psi(k_3)  
\endalign
$$
and 
$$
\align
e^{it\Delta_N}& \[|\psi(\bullet)|^2\(\psi(\bullet+{1\over N}) 
+\psi(\bullet -{1\over N})\)\]_N' (x)\\
= \sum\limits_{|k_i|\leq m}&  e^{it\Delta(N,p)} 
e^{2\pi i p(k_1+k_3- k_2)x} 2\cos{2\pi k_3\over N} 
2\pi i p(k_1+k_3-k_2) \psih(k_1)\psihb(k_2) \psih(k_3), 
\endalign
$$
where $p(k)\equiv k\;\;(\mod\; N)$ and $|p(k)|\leq m$. 

Now,  
$$
\align
&\left|\partial_x e^{it\Delta_N}\[|\psi(\bullet)|^2
\(\psi(\bullet+{1\over N}) + \psi(\bullet -{1\over N})\)\]_N'(x)\right|^2\\
=&\sum\limits_{|k_i|\leq m} e^{it[\Delta(N,p)-\Delta(N,p')]}
 e^{2\pi i x (p(k_1+k_3-k_2)- p(k_4+k_6-k_5))} 2\cos{2\pi k_3\over N} 
2\cos{2\pi k_6\over N}\\
& \quad \quad \times 4\pi^2 p(k_1+k_3-k_2) p'(k_1+k_3-k_5)\psih(k_1) 
\psihb(k_2) \psih(k_3) \psihb(k_4) \psih(k_5) \psihb(k_6).
\endalign
$$ 
Estimating
$$
\phantom{ooo}
E_{\gamma^N} \bigl|\hdots\text{ditto}\hdots\bigr|^{2n}=\bigl|E_{\gamma^N}
|\hdots\text{ditto}\hdots |^{2n}\bigr| 
$$
$$
= | \sum\limits_{j=1}^{n} 
\prod e^{it[\Delta(N,p^j)-\Delta(N,p'{}^j)]}
e^{2\pi i x( p(k_1^j-k_3^j-k_2)-p(k_4^j +k_6^j-k_5^j))}
2\cos{2\pi k_3^j\over N}2 \cos{2\pi k_6^j \over N} 
$$

$$
\align
&\times 
 E_{\gamma^N} \prod 4\pi^2p\,(k_1^j+k_3^j-k_2^j)p'(k_4^j+k_6^j-k_5^j)
\psih(k_1^j)\psihb(k_2^j)\psih(k_3^j)\\
&\phantom{ooooooooooooooooooooooooooooooo}\times\psihb(k_4^j)
\psi(k_5^j)\psihb(k_6^j)|\\
&\leq \sum\limits_{j=1}^{n} E_{\gamma^N}\prod\limits_{|k_i^j|\leq m} 
4\pi^2 |p(k_1^j+k_3^j-k_2^j)||p(k_4^j+k_6^j-k_5^j)\\
&\phantom{ooooooooooooooooooooooooooooo}
\times \psih(k_1^j)\psihb(k_2^j)\psih(k_3^j) \psihb(k_4^j)\psih(k_5^j) 
\psihb(k_6^j).
\endalign
$$
Using the inequalities
$$
|p(k_1+k_3-k_2)|\leq |k_1+k_2-k_3|\leq (1+|k_1|)(1+|k_2|)(1+|k_3|),
$$
and
$$
(1+ |k|)^2\leq 2(1+|k|^2), 
$$
we overestimate the last sum by 
$$
\leq c_{27} \sum\limits_{|p_i|\leq m} E_{\gamma^N}(1+p_1^2)|
\psih(k_1)|^2\hdots \quad \quad \quad \quad \hdots (1+p_1^2)|\psih(p_{3n})|^2.
$$
The worst term 
$$
\sum E_{\gamma^N} p_1^2|\psih(p_1)|^2\hdots\quad \hdots p_{3n}^2|\psih(3n)|^2.
$$
can be overestimated by 
$$
\therefore \leq E_{\gamma^N} |\psi'(x)|^{2\times 3n}\leq C(n,N)
$$
due to $ak^4 + 1\leq \sigma_N^2(k)$.
\qed
\enddemo 

\subhead 11. Identification of measure $d\M_K$\endsubhead This section  
shows that the measure $d\M_K$ is supported on the solutions of the 
NLS flow.

Multiplying both parts of the original equation
$$
i\psi^{\bullet}=-\psi''+2|\psi|^2 \psi
$$
by the test function $f(x,t),\; (x,t)\in \T\times(0,T]$ and integrating  
produces 
$$
\int_{0}^{1}\int_{-\infty}^{+\infty} dx\, dt\[i\psi f^{\bullet} 
-\psi f''+|\psi|^2 2 \psi f\]=0.
$$
\proclaim{Lemma 12} In the statistical ensemble $d\M_K$ 
$$
E_{\M_K}\left|\int_{0}^{1}\int_{-\infty}^{+\infty} dx dt \[i\psi f^{\bullet} - 
\psi f'' +|\psi|^2 2\psi f\]\right|^2=0.
$$
\endproclaim
\demo\nofrills{Proof.\usualspace} The equation for the AL flow on $\M_N$ is  
$$
i\psi^{\bullet}(x,t)= -\Delta_N\psi(x,t) +|\psi(x,t)|^2(\psi(x+{1\over N},t)
+\psi(x-{1\over N},t)),  
$$
where $\Delta_N\psi(x)=N^2\(\psi(x+{1\over N})+\psi(x-{1\over N})-2\psi(x)\)$ 
and $x\in \T_N$. We multiply by the test function $f(x,t)$ 
and integrate over the variables  $t$ and $x$.
$$
\align
\int_{0}^{1}\int_{-\infty}^{+\infty} dx\,dt \, \delta_N(x) [& i\psi(x,t)
f^{\bullet} -\psi(x,t)\Delta_N f \\
+& |\psi(x,t)|^2\(\psi(x+{1\over N},t) +
\psi(x-{1\over N},t)\)f]=0,
\endalign
$$
in which  $\delta_N(x)={1\over N} \sum\limits_{x_0\in \T_N} \delta(x-x_0).
$
Now 
$$
\align
& E_{\M_K}\left|\int\int dx\,dt \[i\psi f^{\bullet}-\psi f''+|\psi|^2 2\psi 
f\] \right|^2\\
= \lim_{N\rightarrow \infty} & E_{\M_K^N} |...\; ditto\; ...|^2\\
= \lim_{N\rightarrow \infty} & E_{\M_K^N} |\int\int dx\,dt \; i\psi f^{\bullet} 
(1- \delta_N(x))- \int\int dx\,dt \;\psi  (\partial_{xx}^2-\delta_N\Delta_N)f\\ 
&+   \int\int dx\,dt  \[|\psi|^22\psi -|\psi|^2 \(\psi(x+{1\over N}) +
\psi(x-{1\over N})\) \delta_N(x)\]f|^2\\
\leq  \lim_{N\rightarrow \infty} c& E_{\M_K^N} \left|\int\int dx \,dt \psi 
f^{\bullet} (1-\delta_N)\right|^2 \\
+ & \lim_{N\rightarrow \infty} c E_{\M_K^N} \left|\int\int dx dt \psi 
(\delta_{xx}^2-\delta_N \Delta_N)f\right|^2 
\endalign
$$
$$
+\lim_{N\rightarrow \infty} c  E_{\M_K^N} \left|\int \int dx\,dt \[|\psi|^22 
\psi -
|\psi|^2\(\psi(x+{1\over N}) + \psi(x-{1\over N})\) \delta_N(x)\] f\right|^2.
$$

The first term  overestimated by 
$$
\align
\therefore  \leq &|f^{\bullet}|_{\infty} E_{\M_K^N} \(\int_0^1\int_0^T 
dx\,dt |\psi(x,t) - \psi(k/N,t)|dx\)^2\\
 \leq & |f^{\bullet}|_{\infty} T \int_0^T E_{\M_K^N} \( \int_0^1|\psi(x,t)-
\psi(k/N,t)|dx\)^2 \\
 \leq & |f^{\bullet}|_{\infty} T^2 \int_0^1 dx E_{\M_K^N} 
|\psi(x,t) - \psi(k/N,t)|^2\\
=& |f^{\bullet}|_{\infty} T^2 N \int^{1/N}_0 E_{\M_K^N}|\psi(x)-\psi(0)|^2 dx\\
\leq & |f^{\bullet}|_{\infty} T^2 \underset0\leq x\leq 1/N\to\sup E_{\M_K^N}
|\psi(x)- \psi(0)|^2=o(1),
\endalign
$$
as $N\rightarrow \infty$ 
due to the stochastic continuity of the random field $\psi$. Estimates for the 
remaining terms can be obtained along the same lines.   
\qed
\enddemo

Due to the results of \cite{MCV1, B2} the initial data $\psi(x,t)$ determines 
the flow $\psi(\bullet,t)$ for all $t$. In probabilistic language this means 
that  $\psi(\bullet,t)$ is measurable with respect to the field generated 
by $\psi(\bullet,0)$.

\bye